\documentclass[onecolumn]{aastex6}

\renewcommand{\vec}[1]{\mathbf{#1}}

\begin{document}
\title{Extension of the MURaM  radiative MHD code for coronal simulations}
\shorttitle{Extension of the MURaM code}

\author{M. Rempel\altaffilmark{1}}

\shortauthors{Rempel}

\altaffiltext{1}{High Altitude Observatory,
    NCAR, P.O. Box 3000, Boulder, Colorado 80307, USA}

\email{rempel@ucar.edu}

\begin{abstract}
We present a new version of the MURaM radiative MHD code that allows for simulations spanning from the upper convection zone into the solar corona.
We implemented the relevant coronal physics in terms of optically thin radiative loss, field aligned heat conduction and an equilibrium ionization equation of
state. We artificially limit the coronal Alfv{\'e}n and heat conduction speeds to computationally manageable values using an approximation 
to semi-relativistic MHD with an artificially reduced speed of light (Boris correction).
We present example solutions ranging from quiet to active Sun in order to verify the validity of our approach. We quantify the role of numerical diffusivity for the effective coronal heating.
We find that the (numerical) magnetic Prandtl number determines the ratio of resistive to viscous heating and that owing to the very large magnetic Prandtl number of the solar corona, heating 
is expected to happen predominantly through viscous dissipation. We find that reasonable solutions can be obtained 
with values of the reduced speed of light just marginally larger than the maximum sound speed. Overall this leads to a fully explicit code that can compute the time 
evolution of the solar corona in response to photospheric driving using numerical time steps not much smaller than $0.1$ seconds. Numerical simulations of the coronal response
to flux emergence covering  a time span of a few days are well within reach using this approach.
\end{abstract}


\received{}
\accepted{}

\maketitle
\section{Introduction}
Comprehensive numerical simulations of the solar corona have been developed by a few teams in the past decade \citep{Gudiksen:Nordlund:2002,Gudiksen:Nordlund:2005a,
Gudiksen:Nordlund:2005b,Abbett:2007,Gudiksen:etal:2011,Bingert:Peter:2011,Bingert:Peter:2013,Bourdin:etal:2013:obs_driven,Chen:etal:2014}.
These simulations are often referred to as ``realistic'' in the sense that they include the relevant macro-physics in terms of MHD, field aligned heat conduction, optically thin 
radiative loss, and a solar mixture equation of state including partial ionization effects. The heating of the corona is implicitly handled through MHD (upward directed Poynting flux above photosphere) in combination with
either explicit or implicit (i.e. numerical) magnetic and viscous diffusivities, which can be considered as a numerical representation of the heating due to braiding of field lines
as first suggested by \citet{Parker:1972:topological_heating,Parker:1983:sheets}, see also the review by \citet{Klimchuk:2006:review} for a general overview. It was demonstrated by 
\citet{Gudiksen:Nordlund:2002,Gudiksen:Nordlund:2005b,Gudiksen:Nordlund:2005a} that
this process is sufficient to maintain the corona at MK temperatures. While the treatment of energy dissipation is not realistic (i.e. these models do not include the correct
micro-physics) it has been found by \citet{Peter:etal:2004,Peter:etal:2006} that these models do compare remarkably well with observations in a statistical sense. A later analysis by \citet{Bingert:Peter:2013}
showed that the statistical properties of the energy deposition agree with the predictions of the nanoflare model \citep{Parker:1988:nanoflare}.
While the models of  \citet{Bingert:Peter:2011,Bingert:Peter:2013,Bourdin:etal:2013:obs_driven,Chen:etal:2014} do not include the upper convection zone and are driven by a boundary condition
either taken from observations or other numerical simulations, the model of \citet{Gudiksen:etal:2011} does self-consistently treat the coupling from the upper convection zone into
the solar corona, including also realistic physics for the photosphere (3D radiative transfer) and chromosphere including non-local thermal equilibrium (NLTE) physics. The coupling from the upper convection zone into the solar corona
was also addressed by \citet{Abbett:2007}, although their model did not use radiative transfer in the photosphere and relied in addition on empirical heating terms since the braiding 
of magnetic field lines by photospheric motions turned out to be insufficient at least in the quiet Sun setup they considered. The use of empirical heating descriptions is more common in codes
that aim at modeling the larger scale corona \citep[see, e.g.][]{Mok:etal:2005,Mok:etal:2008,vdHolst:etal:2014:AWSoM}, where the resolution is in general insufficient to capture the processes involved in coronal heating directly. 

In this paper we present a new version of the MURaM radiative MHD code \citep{Voegler:etal:2005,Rempel:2014:SSD}, in which we implement a treatment of the corona along the lines of the
above mentioned ``realistic'' simulations. Unlike \citet{Gudiksen:etal:2011} we do not implement at this time a realistic treatment of the chromosphere, i.e. our ``chromosphere'' is treated
assuming local thermal equilibrium LTE. Our main emphasis is on introducing efficient ways to deal with the two most stringent numerical  time step constraints that are encountered in direct simulations of the
solar corona: high Alfv{\'e}n velocities, which might even exceed the speed of light in the classical approximation, and severe time step constraints from field aligned heat conduction.
In this paper we explore the potential of the so called Boris correction \citep{Boris:1970:BC} to deal with the high Alfv{\'e}n velocities (semi-relativistic MHD with an artificially reduced speed of
light), and apply a conceptually similar approach also to heat conduction. 

The paper is organized as follows. In Section \ref{sec:num} we summarize the approximations used and present the full set of equations solved. In Section \ref{sec:results} we test the
validity of the approximations for simulation of the solar corona by considering four different setups (quiet Sun, open flux, coronal arcade and active region). Section \ref{sec:heating} presents
an analysis of the time scales that govern coronal energy transport and release. We briefly discuss numerical efficiency
in Section \ref{sec:efficiency} and present our conclusion in Section \ref{sec:concl}.

\section{Numerical approach}
\label{sec:num}
\subsection{Semi-relativistic MHD (Boris correction)}
\label{Sec:Boris}
In an active region corona the Alfv{\'e}n velocity $v_{\rm A}=\vert\vec{B}\vert/\sqrt{4\pi\varrho}$ can reach values exceeding $100,000$~km~s$^{-1}$ or even the speed of light. While the latter is a consequence of
using non-relativistic equations even values beyond $10,000$ km~s$^{-1}$ can impose stringent numerical time step constraints that make simulations covering long
(several day) time scales very expensive. A similar problem was already encountered in Sunspot simulations and was avoided in photospheric simulations by simply
reducing the strength of the Lorentz force \citep{Rempel:etal:2009}. That approach was also adopted for coronal simulations by \citet{Chen:Peter:2015}. There are two significant 
drawbacks of this approach. A reduction of the Lorentz force perturbs the force balance and may lead to unphysical solutions. In addition it leads to an energetic
inconsistency between the momentum and induction equation that can be a concern for simulations that attempt to address coronal heating. Both limitations can be
minimized by selecting a sufficiently high cutoff for the Alfv{\'e}n velocity, typically $V_{A\,{\rm max}} > 10\,C_S$ ($C_S$: speed of sound) in order to ensure that the system remains in a low-$\beta$
state even with the artificially reduced magnetic pressure.

An alternate way of limiting the Alfv{\'e}n velocity was proposed by \citet{Boris:1970:BC}. The approach is based on semi-relativistic MHD with an artificially reduced speed of light and has been widely
used for numerical simulations of planetary magnetospheres \citep[see, e.g.][]{Gombosi:etal:2002:SR,Lyon:etal:2004:LFM}. The semi-relativistic treatment keeps the displacement current in the Maxwell equations,
but neglects all other relativistic terms, i.e. this approach is valid in the regime $v\ll v_A\sim c$, where $c$ denotes the (artificially reduced) speed of light. With the Maxwell equation 
\begin{equation}
	\nabla\times\vec{B}=\frac{4\pi}{c}\vec{j}+\frac{1}{c}\frac{\partial \vec{E}}{\partial t}
\end{equation}
we can rewrite the Lorentz force as
\begin{eqnarray}
	\frac{1}{c}\vec{j}\times\vec{B}&=&\frac{1}{4\pi}(\nabla\times\vec{B})\times\vec{B}+\frac{1}{4\pi c}\vec{B}\times\frac{\partial \vec{E}}{\partial t}	 \nonumber \\
		&=&\frac{1}{4\pi}(\nabla\times\vec{B})\times\vec{B}+\frac{1}{4\pi}(\nabla\times\vec{E})\times\vec{E}-\frac{\partial}{\partial t}\frac{\vec{E}\times\vec{B}}{4\pi c}
\end{eqnarray}
Inserting this expression into the momentum equation leads to
\begin{eqnarray}
	\frac{\partial}{\partial t}&&\left(\frac{\vec{E}\times\vec{B}}{4\pi c}+\varrho\vec{v}\right)+\nabla\cdot(\varrho\vec{v}\vec{v}+\mathcal{I}p)=\varrho\vec{g}
		+\frac{1}{4\pi}(\nabla\times\vec{B})\times\vec{B}+\frac{1}{4\pi}(\nabla\times\vec{E})\times\vec{E}\label{eq:momentum-boris}
\end{eqnarray}
Here $\varrho$ denotes the mass density, $\vec{v}$ the velocity, $\vec{B}$ the magnetic field, $\vec{j}$ the electric current and $\vec{E}$ the electric field given by
$\vec{E}=-\vec{v}/c\times\vec{B}$ with the speed of light $c$. Inserting the expression for $\vec{E}$ into Eq. (\ref{eq:momentum-boris}) leads to an enhancement of the 
effective inertia perpendicular to field lines by a factor of
$1+{v_A^2/ c^2}$, where $v_A={\vert\vec{B}\vert/\sqrt{4\pi\varrho}}$ denotes the non-relativistic Alfv{\'e}n velocity. This leads to an asymptotic limit of the relativistic Alfv{\'e}n
velocity by $c$. The basic idea behind the Boris correction is to use the minimal amount of correction terms needed for limiting $v_A$ and artificially reduce $c$ to a desired cutoff
velocity. Typically the electric stress term on the right hand side is neglected, since only the inertia enhancement leads to an effective reduction of the wave speed.

Unlike the approach of reducing the Lorentz force, the correction term only appears in the time derivative, i.e.
stationary solutions of the system are not affected. Furthermore the system remains energetically consistent and does conserve energy, provided that the energy of the electric field
is included. This approach adds the correction term $ \frac{1}{4\pi c}\vec{B}\times\frac{\partial \vec{E}}{\partial t}	$ in the momentum equation. The work done by this term us given by
\begin{equation}
	\frac{1}{4\pi c}\vec{v}\cdot\left(\vec{B}\times\frac{\partial \vec{E}}{\partial t}\right)=-\frac{\partial}{\partial t}\frac{\vec{E}^2}{8\pi}\,,	\label{eq:energy-boris}
\end{equation}
i.e. it describes the transfer to the electric energy reservoir. Energy that is temporarily ``parked'' in this reservoir is not lost from the MHD system since $E$ is bounded by ${v\over c} B$
and cannot grow indefinitely. For quasi-stationary solutions the energy transfers introduced by the correction term have to average out to zero. 

\subsection{Hyperbolic heat conduction}
\begin{figure*}
  	\centering
   	\resizebox{0.95\hsize}{!}{\includegraphics{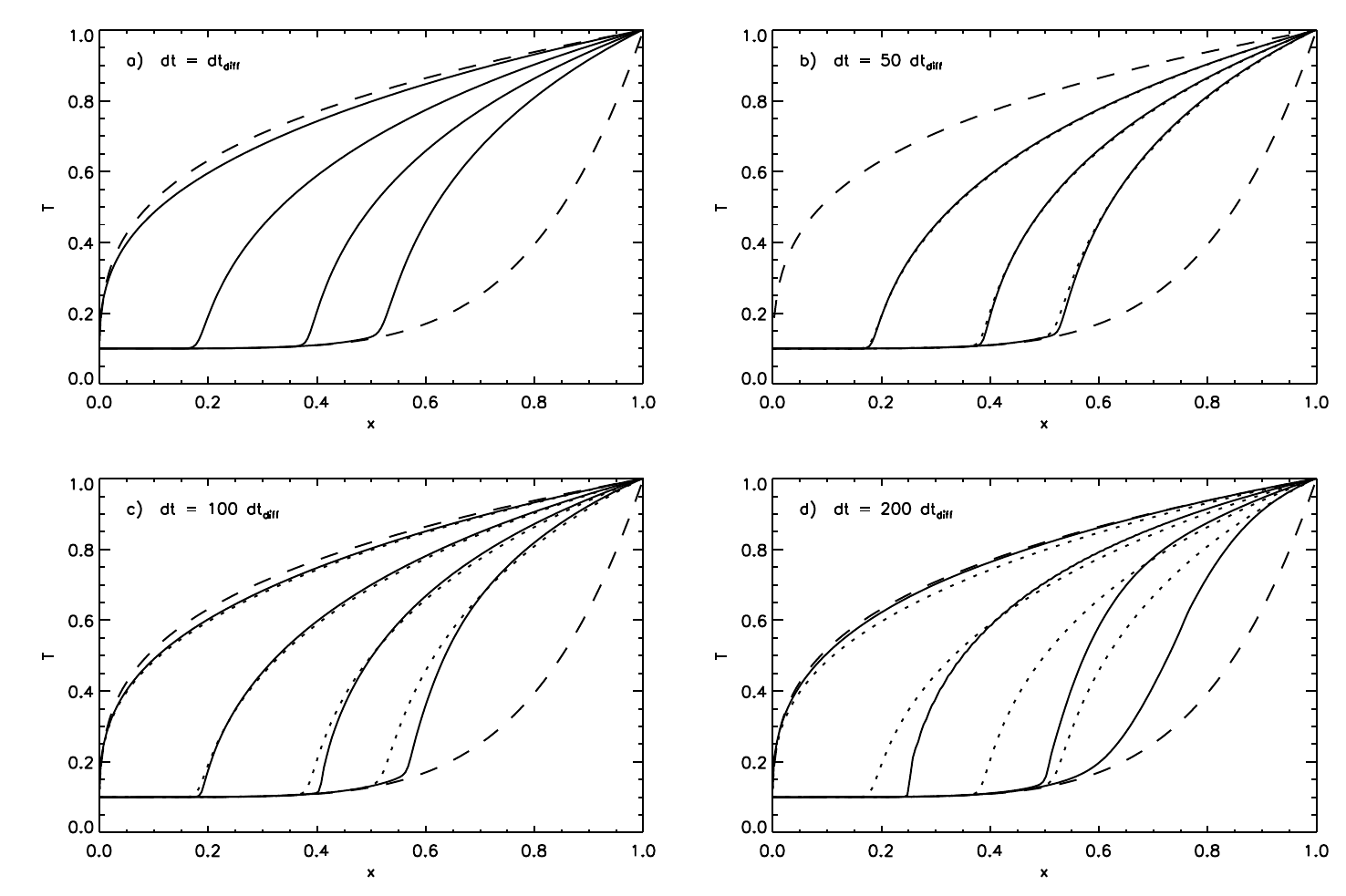}}
   	\caption{1D heat conduction test. The dashed lines indicate the initial state and the asymptotic solution. Panel a): Solution of parabolic heat conduction equation for the times
   	$0.125$, $0.25$, $0.5$, and $1$. Panels b,c,d): Solution of hyperbolic heat conduction equation integrated with $50$, $100$, and $200$ $\Delta t_{\rm diff}$, the reference
   	solution from panel a) is indicated by dotted lines.}
   	\label{fig:1}
\end{figure*}
Under typical coronal conditions the numerical treatment of heat conduction is numerically challenging due to the stringent time step constraint in an explicit treatment. Possible (in part
still expensive) solutions are implicit treatment or super time-stepping schemes such as \citet{Meyer:2012:super-timestepping}. Here we explore a different approach that was used in
the context of galactic cosmic ray transport by \citet{Snodin:etal:2006:non-fickian}. They suggested to use a non-fickian (hyperbolic) diffusion equation, which naturally introduces a maximum
signal propagation velocity due to its hyperbolic character. Formally hyperbolic transport equations for the turbulent transport of passive scalars follow from the minimal $\tau$-approximation 
\citep{Blackman:Field:2003:non-fickian} and have have been verified in direct numerical simulations \citep{Brandenburg:etal:2004:non-fickian}. Non-fickian transport equations are also found
when higher terms are considered in derivations that start with Boltzman-equation \citep[see, e.g.][]{Schunk:1975,Gombosi:1993:telegraph,Lie-Svendsen:etal:2001}. Similar to our treatment of the Alfv{\'e}n
velocity in Section \ref{Sec:Boris}, we use also here non-Fickian (hyperbolic) heat conduction as a physics inspired numerical device to circumvent stringent timestep constraints, which of course has
to be carefully tested similar to the Boris correction for typical coronal conditions. In order to illustrate the approach, we consider first the following simplified 1D system:
\begin{eqnarray}
	\tau\frac{\partial q}{\partial t}+q&=&-\kappa\frac{\partial T}{\partial x}\\
	\frac{\partial T}{\partial t}&=&-\frac{\partial q}{\partial x}
\end{eqnarray}
For $\tau=0$ this sytem is identical to parabolic heat conduction, for $\tau>0$ it corresponds to solving a wave equation for $T$ of the form
(we assume here for simplicity that $\tau$ and $\kappa$ are constant):
\begin{equation}
	\frac{\partial^2 T}{\partial t^2}+\frac{1}{\tau}\frac{\partial T}{\partial t}-c^2\frac{\partial^2 T}{\partial x^2}=0\,
\end{equation}
where the wave propagation speed is given by $c=\sqrt{\kappa/\tau}$. While $\kappa$ is in general given, we choose
$\tau$ such that the maximum wave speed in the hyperbolic heat conduction equation is comparable to the maximum wave
speed of the MHD system with reduced Alfv{\'e}n velocity as described above $(c=f_{\rm CFL}\Delta x_{\rm min}/\Delta t)$. This leads 
to a relation 
\begin{equation}
	\tau = \frac{\Delta t^2 \kappa}{f_{\rm CFL}^2\Delta x_{\rm min}^2}\;,
\end{equation}
Using  $\Delta t_{\rm diff}=f_{\rm CFL}\,\Delta x_{\rm min}^2/(2 \kappa)$ we can write
\begin{equation}
	\tau = \frac{1}{2\,f_{\rm CFL}}\frac{\Delta t^2}{\Delta t_{\rm diff}};,
\end{equation}
Neglecting factors of order unity we can use the expression to estimate the speedup $\alpha$ that is possible with this approach
as
\begin{equation}
	\alpha \sim \frac{\Delta t}{\Delta t_{\rm diff}}\sim \sqrt{\frac{\tau}{\Delta t_{\rm diff}}}
\end{equation}
In order to maintain a solution that does not differ too much from the solution of the parabolic heat conduction equation $\tau$
should remain shorter than a typical time scale of the system. If we assume that the system is dominated by heat conduction,
that time scale itself is given by $L^2/\kappa$, where $L$ is a typical length scale, i.e. the length of a coronal loop in the case of
the solar corona. If we require $\tau \le L^2/\kappa$ , the maximum acceptable speedup is simply given by the numerical resolution
\begin{equation}
	\alpha\le\frac{L}{\Delta x_{\rm min}}
\end{equation}
For a typical coronal setup with $L$ on the order of a few $10$~Mm and $\Delta x_{\rm min}$ around $100$~km speedups on the order
of a few $100$ can be possible.

We illustrate the hyperbolic heat conduction approach with the help of a one-dimensional solution of the heat conduction equation in
a domain of the size $0\le x \le 1$ and the temperature boundary conditions $T(0)=0.1$ and $T(1)=1$. We assume a Spitzer like heat
conductivity by choosing $\kappa=T^{2.5}$. Under these assumptions the asymptotic stationary solution is given by
\begin{equation}
	T_s(x)=\left[0.1^{3.5}+(1-0.1^{3.5})x\right]^{2/7}
\end{equation}
We choses an initial non-equilibrium solution of the form 
\begin{equation}
	T_i(x)=0.1+0.9 x^5
\end{equation} 
and follow the time evolution towards the asymptotic solution. In order to simulate a setup comparable to a typical coronal simulation we use
a value of $\Delta x = 0.005$, i.e. a resolution of $250$ gridpoints along a structure. This leads to a diffusive time step 
$\Delta t_{\rm diff} = f_{\rm CFL}\cdot{1\over2} \Delta x^2=f_{\rm CFL}\cdot 10^{-5}$ ($\kappa_{\rm max}=1$), while the asymptotic solution is reached 
for times of order unity. The reference solution computed with $\tau=0$ and $f_{\rm CFL}=0.8$ is presented in Figure \ref{fig:1}a) for the
times $0.125$, $0.25$, $0.5$ and $1$. Dashed lines indicate $T_i$ and $T_s$. In panels (b) - (d) we present the corresponding solutions
computed with the hyperbolic heat conduction equation using $\tau = \Delta t^2 T^{2.5}/(f_{\rm CFL}^2\Delta x^2)$ for the values $\Delta t=$
$50$, $100$ and $200$ $\Delta t_{\rm diff}$. The maximum $\tau$ values for these choices are $0.015625$, $0.0625$, and $0.25$, respectively.
For the speedup of $50$ the solution is very similar to the reference solution for all times steps shown. In the case with a speedup of $100$ 
clear differences exist for time smaller than $0.25$, in the case with speedup of $200$ only the final time of $1$ is similar to the reference
solution, while earlier times show a clear slowdown of the propagation of the heat front as expected. From this experiment we can conclude that
$\tau$ should remain about a factor of $4$ shorter than the time scale of interest in order to maintain a solution close to the reference solution.
We note that in all cases the asymptotic solution is identical. If the main focus of a simulation is mostly the average energy balance of the corona
with less emphasis on the time evolution of individual features even larger speedups can be tolerated. We discuss the limitations of this
approach for the solar corona further in Section \ref{sec:conduction}.

For the implementation in the 3-dimensional coronal MHD code we have to account for field aligned heat conduction, which is achieved by the following set of equations:
\begin{eqnarray}
	\frac{\partial q}{\partial t}&=&\frac{1}{\tau}\left(-f_{\rm sat}\sigma T^{5\over2}(\hat{\vec{b}}\cdot\nabla)T-q\right)\label{eq:flux}\\
	\frac{E_{\rm int}}{\partial t}&=&[\ldots]-\nabla\cdot (q\hat{\vec{b}})\\
	\tau&=&\left(f_{\rm CFL}{\Delta x_{\rm min}\over\Delta t}-|v|\right)^{-2}\frac{f_{\rm sat}\sigma T^{7\over2}}{E_{\rm int}}\label{eq:tau}
\end{eqnarray}
Here $\hat{\vec{b}}=\vec{B}/|\vec{B}|$ is the unit vector in the direction of the field. We use here a value of $10^{-6}\mbox{erg}$ $\mbox{cm}^{-1}\mbox{s}^{-1} \mbox{K}^{-{7\over 2}}$  for the constant $\sigma$ of
the Spitzer heat conductivity \citep{Spitzer:1962}. The pre-factor $f_{\rm sat}$ considers the saturation of the conductive heatflux following \citet{Fisher:etal:1985:flare,Meyer:2012:super-timestepping}. We use here
\begin{equation}
	f_{\rm sat}=\left(1+\frac{\vert \sigma T^{5\over2}(\hat{\vec{b}}\cdot\nabla)T \vert}{1.5\varrho C_S^3}\right)^{-1}\,\label{Eq:saturation}
\end{equation}
where $C_S=\sqrt{\gamma p/\varrho}$ denotes the speed of sound. Including the factor $f_{\rm sat}$ in Eq. (\ref{eq:tau}) leads in general to a significantly less smooth profile of $\tau$,
but has the advantage that the values of $\tau$ are generally lower, i.e. the approximation is better suited to capture fast changes. In the simulations presented here we used
a more conservative approach and excluded the factor $f_{\rm sat}$ in Eq. (\ref{eq:tau}), while keeping it in Eq. (\ref{eq:flux}). More tests including a more dynamic flare setup have shown since then that 
including  $f_{\rm sat}$ also in Eq. (\ref{eq:tau}) does not introduce any artifacts in the quantity $q$. In the setups considered in this paper the heat flux barely reaches saturation values.
In the expression for $\tau$ we use the quantity $f_{\rm CFL}{\Delta x_{\rm min}\over\Delta t}-|v|$ as maximum propagation speed in 
order to avoid violations of CFL condition in regions where conductive and advective transport of heat add up. Since we integrate our system of equations explicitly we have to
impose on $\tau$ a lower limit $\tau_{\rm min}$. For all simulations presented here we use $\tau_{\rm min}=4\Delta t$. 

\begin{figure}
  	\centering
   	\resizebox{0.66\hsize}{!}{\includegraphics{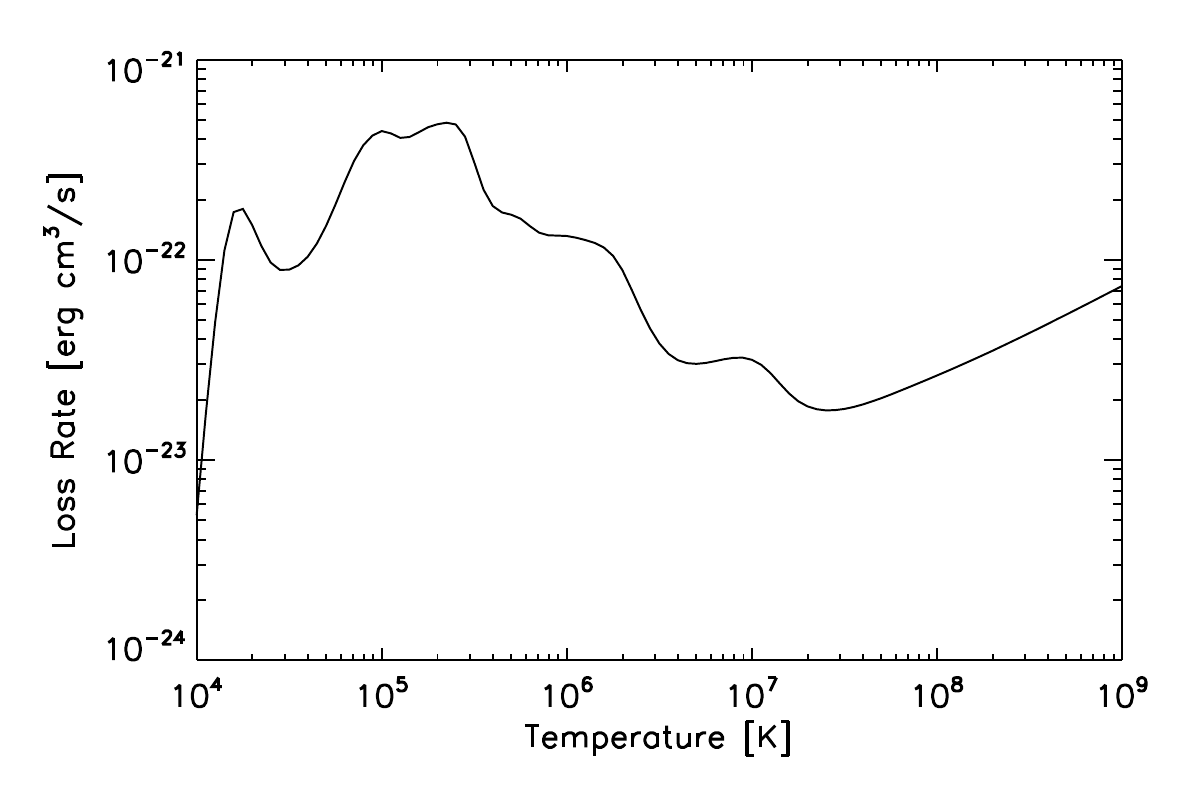}}
   	\caption{Density averaged radiative loss rate $\Lambda(T)$ based on the Chianti photospheric abundances.}
   	\label{fig:2}
\end{figure}

\subsection{Radiative loss}
We use in our model a combination of LTE radiative transfer and tabulated optically thin radiative loss. The radiative transfer module, originally developed by \citet{Voegler:etal:2005} was 
modified to dynamically switch off once the transition region is reached. We compute the position of the transition region on a column-by-column basis by finding the first grid cell with $T>20,000$~K 
going upward from the photosphere. In all grid cells above this position (minus a safety distance of one or two grid cells) we set the opacity and source function to zero in order to avoid contributions
from the corona in the downward directed rays.

Starting with the position of the transition region (we formally make the transition at an optical depth of $10^{-8}$) we use an optically thin radiative loss function of the form
\begin{equation}
	Q_{\rm loss}=-n_e\,n_H\Lambda(T)\;,
\end{equation}
where $n_e$ and $n_H$ are given by (assuming a H/He mixture):
\begin{eqnarray}
	n_e&=&\frac{\varrho}{m_p}\frac{1+X}{2}\\
	n_H&=&\frac{\varrho}{m_p}X
\end{eqnarray}
In the following we use $X=0.7$.
For the function $\Lambda(T)$ we use a tabulated loss function taken from CHIANTI 7 with photospheric abundances \citet{Landi:2012:Chianti7}. Formally Chinati provides in a addition
a weak ($10\%$) density dependence in $\Lambda(T)$ that we suppress by using a density averaged loss function as shown in Figure \ref{fig:2} (we average the loss function in the
range from $10^8$ to $10^{12}$ particles per cm$^3$, the average is performed ${\rm log}\,n_e$ weighted).

\begin{figure}
  	\centering
   	\resizebox{0.66\hsize}{!}{\includegraphics{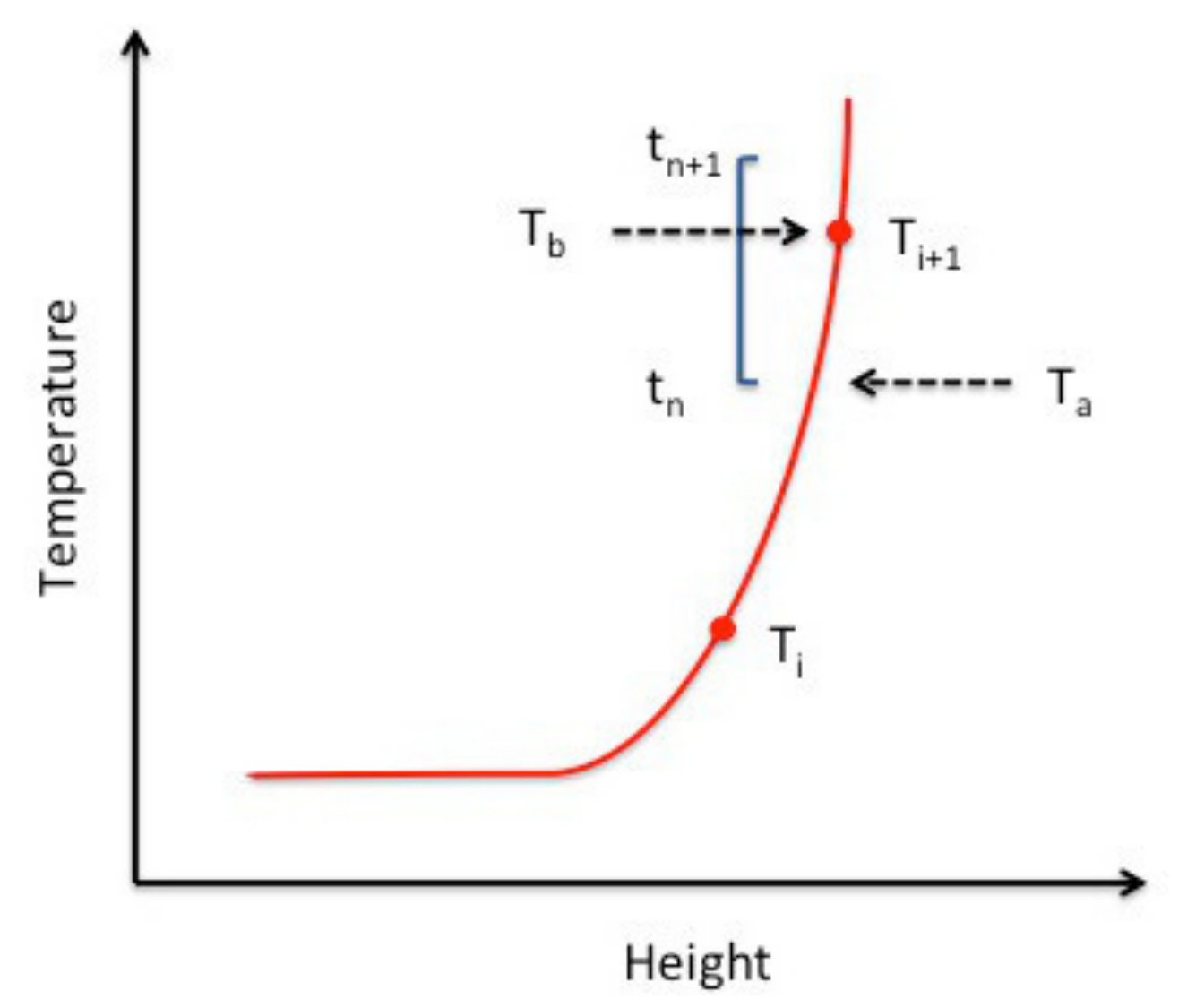}}
   	\caption{Sketch explaining our procedure for computing radiative losses. $T_{i}$ and $T_{i+1}$ are the temperature at two neighboring grid points, $t_n$ and $t_{n+1}$ are sampling
   		points for the tabulated radiative loss function. We consider the overlap between the intervals $[T_i, T_{i+1}]$ and $[t_n,t_{n+1}]$ that is given by
   		$[T_a,T_b]$ to compute a continuous contribution to the radiative loss function as explained in the text.}
   	\label{fig:3}
\end{figure}

A major numerical challenge comes from the fact that the transition region can have a width smaller than a grid spacing, which does lead to an inaccurate
determination of the radiative loss since $Q_{\rm loss}$ peaks at transition region temperatures. Using 1-dimensional hydro models \citet{Bradshaw:Cargill:2013:resolution}
investigated the effect of numerical resolution and found that grid spacings as low as a few km are required in order to properly resolve the width and thermal loss from the
transition region as predicted from Spitzer conductivity. Such grid spacings cannot be achieved in 3-dimensional simulations as presented here. 
Using a lower numerical resolution leads to a significant temperature variation on the scale of the grid, which can both over and underestimate radiative losses.
On the one hand, if the ``sweet spot'' of the transition region happens to fall right on the values found in a grid cell, radiative 
loss is overestimated since the geometric width of the grid cell is too large; on the other hand, if the ``sweet spot''  falls in between two grid cells radiative
loss is underestimated since the transition region is omitted numerically. While both effects cancel out to some degree when considering horizontal averages of the
radiative loss, they can cause a significant artificial variation of $Q_{\rm loss}$ between neighboring grid points and in time when the position of the transition region is moving 
across the simulation grid. In order to minimize this artificial variation and in order
to capture a thickness of the transition region potentially smaller than a gid spacing we use the following scheme that takes into account also information from the
temperature gradient by formally oversampling the solution assuming a linear variation of $\lg T$ and $\lg \varrho$ in-between grid points in the vertical direction.

The radiative loss function is assumed to be tabulated ($\Lambda_n(\lg t_n)$, $n=1 \ldots N_{\rm tab}$).  Rather than checking whether a grid point $\lg T_i$ lies within
a table intervall $[\lg t_n, \lg t_{n+1}]$, we check if there is an overlap between the intervals $[\lg T_i, \lg T_{i+1}]$ and $[\lg t_n,\lg t_{n+1}]$. In the following we assume $\lg T_i < \lg T_{i+1}$
for simplicity. If an overlap is present, we compute the overlap interval
$[\lg T_a,\lg T_b]=[\lg T_i, \lg T_{i+1}]\cap [\lg t_n, \lg t_{n+1}]$. For example if we have the situation (see Figure \ref{fig:3}) $\lg T_i < \lg t_n < \lg T_{i+1} < \lg t_{n+1}$ we have $\lg T_a=\lg t_n$ and $\lg T_b=\lg T_{i+1}$.
Based on the overlap interval we compute the quantities:
\begin{eqnarray}
	f_n&=&\frac{\lg T_b-\lg T_a}{\lg T_{i+1}-\lg T_i}\\
	\lg\bar{T}_n&=&0.5 (\lg T_a+\lg T_b)\\
	x_n&=&\frac{\lg T_{i+1}-\lg\bar{T}_n}{\lg T_{i+1}-\lg T_i}\\
	\lg\bar{\varrho}_n&=&x_n{\lg\varrho}_i+(1-x_n){\lg\varrho}_{i+1}\,.
\end{eqnarray}
The contribution to the total radiative loss function from the interval $[T_i,T_{i+1}]$ is then given by
\begin{eqnarray}
	Q_i&{+\hspace{-0.1cm}=}&\sum_1^{N_{\rm tab}} x_n f_n \left(\frac{\bar{\varrho}_n}{\mu}\right)^2\Lambda(\lg \bar{T}_n)\\
	Q_{i+1}&+\hspace{-0.1cm}=&\sum_1^{N_{\rm tab}} (1-x_n) f_n \left(\frac{\bar{\varrho}_n}{\mu}\right)^2\Lambda(\lg \bar{T}_n)\,
\end{eqnarray}
where $\Lambda(\lg \bar{T}_n)$ is the loss function interpolated in $[\lg t_n,\lg t_{n+1}]$ and $\mu=m_p/\sqrt{0.5(1+X)X}$. Note that $f_n=0$, if there is
no overlap between $[\lg T_i, \lg T_{i+1}]$ and $[\lg t_n, \lg t_{n+1}]$. We use here ``$+\hspace{-0.1cm}=$'' to 
indicate that there are additional contributions to $Q_i$ from the interval $[T_{i-1}, T_{i}]$ and to $Q_{i+1}$ from $[T_{i+1}, T_{i+2}]$
We apply the above procedure in the vertical direction on a column by column basis. Since the transition region can be strongly warped we 
found comparable results when applying the method to the horizontal directions. It is likely that a better accuracy can be obtained by applying the above procedure 
to all 3 grid directions and then considering (a properly weighted) average of the result, but we did not further investigate that possibility.
For the simulations presented here we found that this method leads to an about $25-50\%$ larger radiative loss for a given snapshot compared to the simple point 
by point table lookup procedure. 

We also found the above outlined procedure very useful for extracting the differential emission measure (DEM)
\begin{equation}
	\mbox{DEM}(T)\mbox{d}T=\int_0^{\infty} n_e(T) n_H(T)\mbox{d}s
\end{equation}
from our simulations, where $\mbox{d}s$ could be any direction of interest. The above procedure avoids the problem that certain temperature values (or intervals) might not be 
found in the discrete solution, which would lead to DEMs with ``missing'' data. In Figures \ref{fig:4} to \ref{fig:7} we show the emission mesure (EM) for $\Delta \mbox{lg}(T)=0.1$ 
temperature bins for the vertical and a horizontal direction based on this approach.  

In the above description we assumed a linear variation in $\lg T$ and $\lg \varrho$. Using other functional forms does lead to similar results as long as
they are consistent with a slowly varying pressure across the transition region ($H_p$ is much larger than the width of the transition region), which implies
in leading order a $\varrho\sim 1/T$ relation. For example using a linear variation in $T$ and $p$ and then computing $\varrho$ from the interpolated 
$T$ and $p$ values leads to comparable results for the radiative loss within $5\%$. For extracting DEM information we found that the interpolation in $\lg T$
and $\lg \varrho$ results in spatial distributions of DEMs with fewer artifacts.

\subsection{Set of MHD equations solved}
Overall we solve the following set of equations (we do not explicitly spell out numerical diffusivities, they are described in Section \ref{sec:num_diff}):
\begin{eqnarray}
\frac{\partial \varrho}{\partial t}&=&-\nabla\cdot\left(\varrho \vec{v}\right)\label{eq:cont}\\
\frac{\partial \varrho\vec{v}}{\partial t}&=&-\nabla\cdot(\varrho\vec{v}\vec{v})-\nabla P +\varrho\vec{g}+\vec{F_L}+\vec{F}_{SR}\label{eq:mom}\\
\frac{\partial E_{\rm HD}}{\partial t}&=&-\nabla\cdot\left[\vec{v}\,(E_{\rm HD} + P)+q\hat{\vec{b}}\right]+\varrho\vec{v}\cdot\vec{g}
	+\vec{v}\cdot \vec{F_L}+\vec{v}\cdot \vec{F}_{SR}+Q_{\rm rad}+Q_{\rm loss}\label{eq:ener}\\
\frac{\partial q}{\partial t}&=&\frac{1}{\tau}\left(-f_{\rm sat}\sigma T^{5\over2}(\hat{\vec{b}}\cdot\nabla)T-q\right)\label{eq:cond}\\	
\frac{\partial \vec{B}}{\partial t}&=&\nabla\times\left(\vec{v}\times\vec{B}\right)\label{eq:ind}
\end{eqnarray} 
Here we use the plasma energy $E_{\rm HD}=E_{\rm int}+{1\over 2} \varrho v^2$ instead of the total energy in order to avoid numerical instabilities in low-$\beta$ regions,
while retaining a conservative treatment for the hydrodynamic part of the equations.
The quantity $\tau$ is given by Eq. (\ref{eq:tau}), $Q_{\rm rad}$ denotes radiative heating/cooling from 3D radiative transfer treatment in regions
with an optical depth larger than $10^{-8}$, while $Q_{\rm loss}$ denotes optically thin radiative loss from regions with an optical depth smaller than
$10^{-8}$ as described above.

The semi-relativistic correction $\vec{F}_{SR}$ is given by (see Appendix):
\begin{eqnarray}
	\vec{F}_{SR}&=&-(1-f_A)\left[\mathcal{I}-\hat{\vec{b}}\hat{\vec{b}}\right]\left(-\varrho(\vec{v}\cdot\nabla)\vec{v} -\nabla p+\varrho\vec{g}+\vec{F_L}\right)\;.\label{eq:fsr}
 \end{eqnarray}
The contribution from numerical viscous forces (not shown here) has to be included in the semi-relativistic correction term Eq. \ref{eq:fsr} for consistency.
Viscous forces, along with all other diffusivities in the code, are computed following the approach detailed in \citet[][see, Section 2.1]{Rempel:2014:SSD}.
The numerical scheme computes diffusive fluxes at cell interfaces based on monotonicity constraints. The divergence of the viscous fluxes gives the viscous forces
we considere here.

The Lorentz force $\vec{F_L}$ is computed numerically as
\begin{eqnarray}
	\vec{F_L}&=&f_A\frac{1}{4\pi}\nabla\cdot\left(\vec{B}\vec{B}-\frac{1}{2}I\vec{B}^2\right)+(1-f_A)\frac{1}{4\pi}(\nabla\times\vec{B})\times\vec{B}
\end{eqnarray}	
as a compromise between a conservative treatment in high $\beta$-regions and the avoidance of spurious field aligned
forces in low $\beta$-regions. The factor $f_A$ determines the functional form of the Alfv{\'e}n velocity limitation, we use here the expression
 \begin{equation}
 	f_A=\frac{1}{\sqrt{1+({v_A\over c})^4}}\,
 \end{equation}
 leading to 
 \begin{equation}
 	v_A^2 \longrightarrow \frac{v_A^2}{\sqrt{1+({v_A\over c})^4}}\;.
 \end{equation}	
 in order to have a sharper transition between the uncorrected and corrected regime of the momentum equation.
 
 The $\nabla\cdot\vec{B}$ error is controlled using the hyperbolic $\nabla\cdot\vec{B}$ cleaning approach of \citet{Dedner:etal:2002:divB}.
 
 \subsection{Equation of state}
 We use a combination of the OPAL equation of state \citep{Rogers:opal:1996} in regions with a density larger than $10^{-6}$ g cm$^{-3}$ 
 and an equation of state based on the Uppsala Opacity Package \citet{Gustafsson:etal:1975}, which was kindly provided to us by the BIFROST team
 \citep{Gudiksen:etal:2011}. The two equations of state are smoothly merged in a single table. For corona temperatures outside the table bounds (above 5 MK) we use an 
 ideal equation of state assuming $\gamma=1.65$ and $\mu= 0.62\,m_p$. $\gamma$ and $\mu$ were chosen to smoothly match the table values 
 near the upper bound. Our equation of state tables assume the LTE (local thermodynamic equilibrium) approximation throughout the simulation
 domain, including the chromosphere.

\subsection{Numerical diffusivity}
\label{sec:num_diff}
In the above equations we did not include viscous and resistive terms since we use only numerical diffusivities as described in
\citet{Rempel:2014:SSD}. We use here the same approach with a few minor modifications. As in \citet{Rempel:2015:moat} we set the diffusive numerical flux
of $B_z$ in the $z$-direction to zero at the boundaries and we reduce the numerical diffusivity of $B$ in the direction of $B$ by a factor of $0.2$. These changes
reduce the $\nabla\cdot\vec{B}$ error and prevent a slow drift of the magnetic flux content in the simulation domain, which can be a problem for simulations covering long
time scales.

Since we use in our formulation the plasma energy equation, numerical resistive heating has to be explicitly added in Eq. (\ref{eq:ener}) (it would be implicit
if we would use the total energy equation). The numerical resistive heating is computed as follows (for clarity in the presentation we consider here only
one grid direction $i$): As described in Section 2.1 of \citet{Rempel:2014:SSD} the numerical scheme provides fluxes at cell interfaces $f_{i+{1\over 2}}^m$ for 
each magnetic field component $B^m$ (m=1,2,3). The numerical scheme is dimensionally split, i.e. contributions from the other grid directions are added after 
each other following the same approach. From the diffusive numerical flux $f_{i+{1\over 2}}^m$  we compute the heating at the grid point $i$ as follows:  
\begin{equation}
	Q_i=-\frac{1}{2}\sum_{m=1}^3\left(f_{i-{1\over 2}}^m\frac{B_i^m-B_{i-1}^m}{\Delta x}+f_{i+{1\over 2}}^m\frac{B_{i+1}^m-B_{i}^m}{\Delta x}\right)\label{Eq:Qres}
\end{equation}
The numerical fluxes $f_{i+{1\over 2}}^m$  are formulated in a way to ensure that $Q$ is positive definite (no anti-diffusion or artificial steepening). Since we 
use an energy equation of the sum of internal and kinetic energy the viscous heating is taken care of implicitly through the conservation
of energy. For physical consistency we do add the corresponding viscous energy flux resulting from numerical viscosity in the energy equation.
We use in our analysis viscous heating for diagnostic purposes, it is computed similar to the resistive heating we explained above.

As described in \citet{Rempel:2014:SSD} the formulation of the numerical diffusivity has a free parameter $h$ that influences the (hyper) diffusive behavior of the scheme.
While a setting of $h=0$ leads to a standard second order TVD Lax-Friedrichs scheme, larger values of $h$ concentrate the diffusivity more around monotonicity changes
of the solution. In addition settings of $h>1$ completely disable the diffusivity in regions that are sufficiently smooth, i.e. the solution has to exceed a certain roughness before
diffusivity kicks in. Similar to \citet{Rempel:2014:SSD} we use a setting of $h=2$ in the convection zone and photosphere. In regions with $\varrho < 10^{-11}$~g~cm$^{-3}$ 
or where $v_A$ exceeds our chosen value of $c$ we use a setting of $h=1.25$ for mass, momentum and energy diffusion and a setting of $h=5$ for the magnetic field. The
latter was chosen to numerically emulate a high magnetic Prandtl number regime for the following 2 reasons: (1) Owing to the low-$\beta$ conditions of the corona the magnetic
field is very smooth (except for unavoidable discontinuities) and therefore does not require a large numerical diffusivity for stability. (2) Based on the expressions for resistivity and viscosity 
from \citet{Spitzer:1962}
\begin{eqnarray}
	\eta = 5.2\times 10^{11} \ln\Lambda T^{-1.5} \mbox{cm}^2\mbox{s}^{-1}\label{eq:eta_spitzer}\\
	\nu  = 2.21\times 10^{-15}\frac{T^{2.5}}{\ln\Lambda\;\varrho} \mbox{cm}^2\mbox{s}^{-1}\label{eq:nu_spitzer}
\end{eqnarray}
the magnetic Prandtl number is given by (using $\ln\Lambda=20$):
\begin{equation}
	P_{\rm m}=\frac{\nu}{\eta}=10^{-29}[\mbox{g}\,\mbox{cm}^{-3}\,\mbox{T}^{-4}]\frac{T^4}{\varrho}\;,
\end{equation}
which yields very high values on the order of $10^{10}$ or larger for the solar corona. Owing to the mostly collision less conditions of the solar corona the detailed physical
interpretation of the diffusivities and the resulting $P_{\rm m}$ is non-trivial, however they do provide some guidance for numerical codes, where the use of 
diffusivities (either explicit or implicit/numerical) is unavoidable. We will further discuss the influence of the numerical magnetic Prandtl number in Section \ref{sec:pm}
and provide estimates of the effective numerical diffusivity in Section \ref{sec:num_diff}. 

We enhance the values of numerical diffusivity in the upper most 1.5 Mm of the simulation domain for reasons of numerical stability at the top boundary condition. 
The enhancement is achieved by multiplying the slopes used in the piece-wise linear reconstruction of the scheme by a factor $\zeta(z)$ given by:
\begin{equation}
\zeta(z)=1-\left(\frac{z-(z_{\rm top}-\Delta z)}{\Delta{z}}\right)^2\,{\rm for}\, z\ge z_{\rm top}-\Delta z\,,
\end{equation}
with $\Delta z=1.5$~Mm. Right at the boundary the reconstruction slopes are zero, i.e. the maximum possible diffusivity ${1\over2} C_{\rm max}\Delta x$ is applied in all three
grid directions ($C_{\rm max}$ denotes the maximum characteristic velocity as given in the Appendix in Eq. (\ref{Eq:Cmax})).
The resistive and viscous heating is not added back to the internal energy in this boundary layer.

\subsection{Boundary conditions}
We use periodic boundary conditions in the horizontal direction. For the test cases discussed later our simulation domain reaches from about $8$~Mm
beneath to about $41$~Mm above the photosphere. The bottom boundary conditions are similar to those discussed in 
\citet{Rempel:2014:SSD}. We use here in particular the boundary ``O16b'' described therein, which imposes a symmetric boundary condition on all three mass flux and magnetic field components. 
While the mean gas pressure at the boundary is fixed, pressure perturbations are damped. At the top boundary we impose a potential field extrapolation and use a 
semi-transparent boundary for flows that allows for mass flux crossing the boundary but strongly damps vertical flows to maintain numerical stability
(boundary values in the 1st (2nd) ghost-cell are set to $50\%$ ($25\%$) of the respective upper domain values). The vertical component of the conductive heat flux 
is set to zero at the boundary. Overall this choice of boundary conditions leads to a setup that is energetically (mostly) closed with respect to advective and conductive energy
fluxes. The boundary does not allow for a Poynting flux leaving the domain, which leads to the formation of a thin boundary layer where most of the
remaining energy flux is dumped through a combination of resistive heating and work done by the Lorentz force on the flows. Since we do not consider resistive and viscous heating
in the uppermost $1.5$ Mm of the simulation domain, most of the remaining Poynting flux is essentially lost.

 \subsection{Choosing the maximum allowed Alfv{\'e}n velocity}
 For solving the above set of equations we need to determine a value for $c$, which is in general a compromise between computational cost and the
 desire to capture the underlying physics sufficiently accurately. The latter is not necessarily achieved by simply using the true speed of light, since the
 maximum characteristic velocities also determine the required numerical diffusivity. Using large values of $c$ would lead to an unrealistically diffusive 
 corona, with diffusive time scales that are too short compared to the fixed time scale of photospheric driving. In general it will be required to repeat
 simulations with a few values of $c$ to assess how sensitive results are with respect to a chosen values of $c$. At a minimum $c$ should be larger
 than $C_S$, since there is obviously no benefit of reducing $c$ to lower values. Our system of equations is based on semi-relativistic MHD that is
 only valid if the maximum flow speed remains small compared to $c$. We did not find that the equations become unstable when $v_{\rm max}$ reaches or
 even exceeds $c$, but as a safeguard that situation should be avoided by either choosing a sufficiently high value of $c$ or by imposing an additional
 limit on the maximum flow speed. We explore in the following both possibilities. For practical purposes we found it useful to use an approach where we
 dynamically adjust $c$ through a relation $c=\alpha v_{\rm max}$ with a value of $\alpha >1$. We note that this approach is inconsistent with with 
 semi-relativistic MHD where $c$ has to be a constant, however, our aim is not to compute an accurate solution of semi-relativistic MHD, but rather to
 use semi-relativistic MHD as a device to limit the Alfv{\'e}n velocity. We did not find any artifacts from dynamically adjusting $c$ in the system of equations
 we are solving. We use for all simulations presented here a value of $\alpha=3$. We found that using a lower value of $1.5$ can lead in some situations to a 
 energetic correction term $\vec{v}\cdot \vec{F}_{SR}$ that does not average out to zero. Since we omit some terms of order $v^2/c^2$, the
 relation expressed in Eq. (\ref{eq:energy-boris}) holds only within that order of approximation.
 
 We do find in our simulations some regions with very large values of the advection speed, which can impose a severe time step limit. Typically these regions 
 concern only a few grid points and have a very low density. In the simulations presented here we impose a dynamically adjusted upper limit for the advection
 velocity. We keep track of the number of gridpoints that have velocities within $5\%$ of our imposed limit. If this number exceeds $1000$ ($0.01\%$ of the simulation
 domain)  we increase the limit, if it falls below $500$ we lower the limit. In addition we increase numerical viscosity for velocities within $25\%$ of the limit by setting
 the slopes used for the piece-wise linear reconstruction to zero, i.e. using the maximum possible diffusivity ${1\over2} C_{\rm max}\Delta x$.
 For the setups considered here this results in maximum velocities in the  $100-300$~km~s$^{-1}$ range, while typical RMS velocities are in the $20-40$~km~s$^{-1}$ 
 range.

\section{Results}
\label{sec:results}

\begin{figure*}
  	\centering
   	\resizebox{0.95\hsize}{!}{\includegraphics{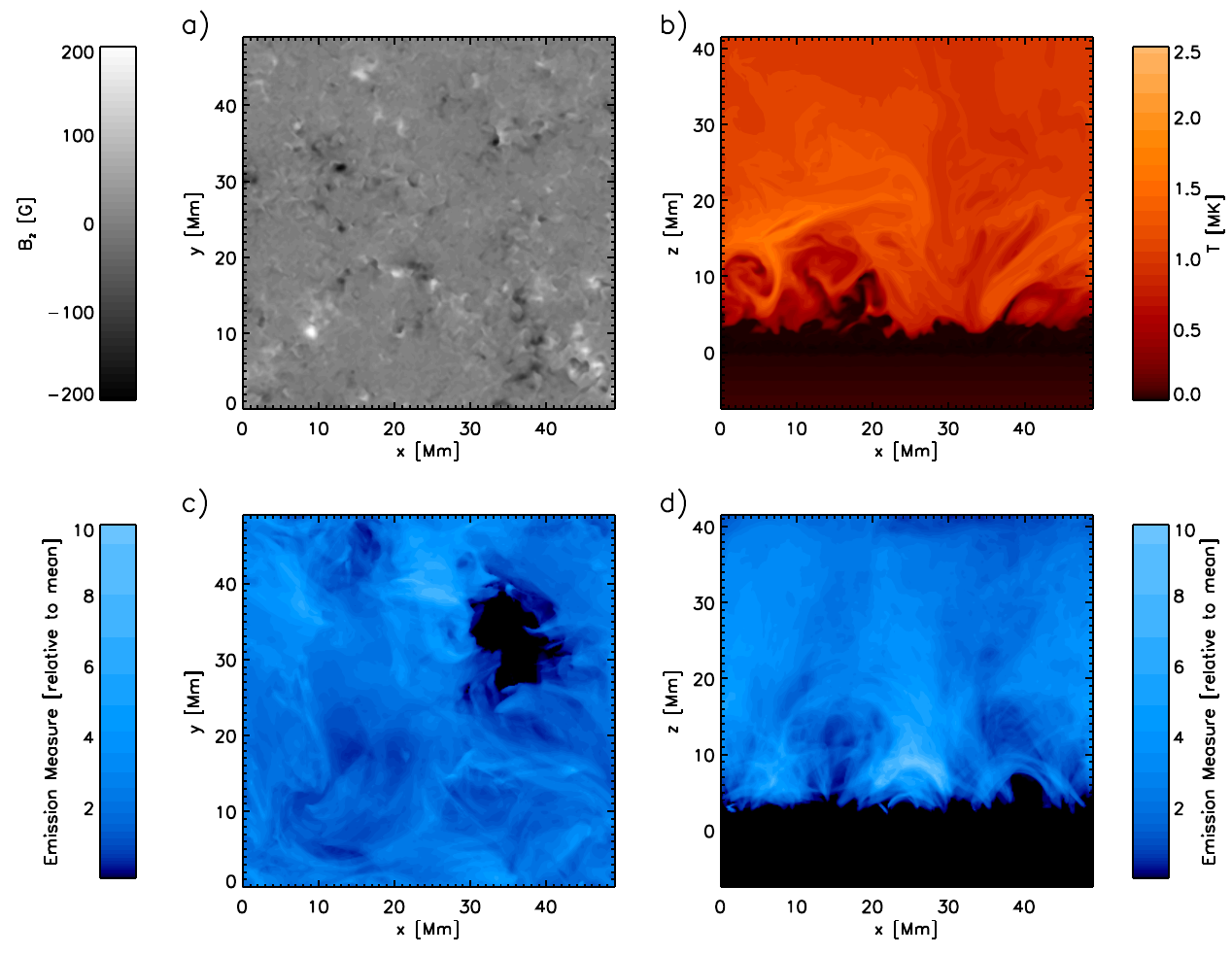}}
   	\caption{Quiet Sun (QS) setup. (a) Vertical magnetic field $700$ km above photosphere; (b) temperature profile on a vertical cut through the simulation domain ($y=24.576$~Mm); 
   		(c) emission measure for a vertical view; (d) emission measure for a side view along the y-axis. The emission measure is presented for all cases for the $\mbox{lg}(T/K)=6.0-6.1$ 
   		temperature interval. The emission in panel (c) is shown on a logarithmic scale ranging from $10^{-2}$ to $10$ times the horizontal mean emission, panel (d) uses the same scale as 
		panel (c). This figure is available as an animation.}
   	\label{fig:4}
\end{figure*}

\begin{figure*}
  	\centering
   	\resizebox{0.95\hsize}{!}{\includegraphics{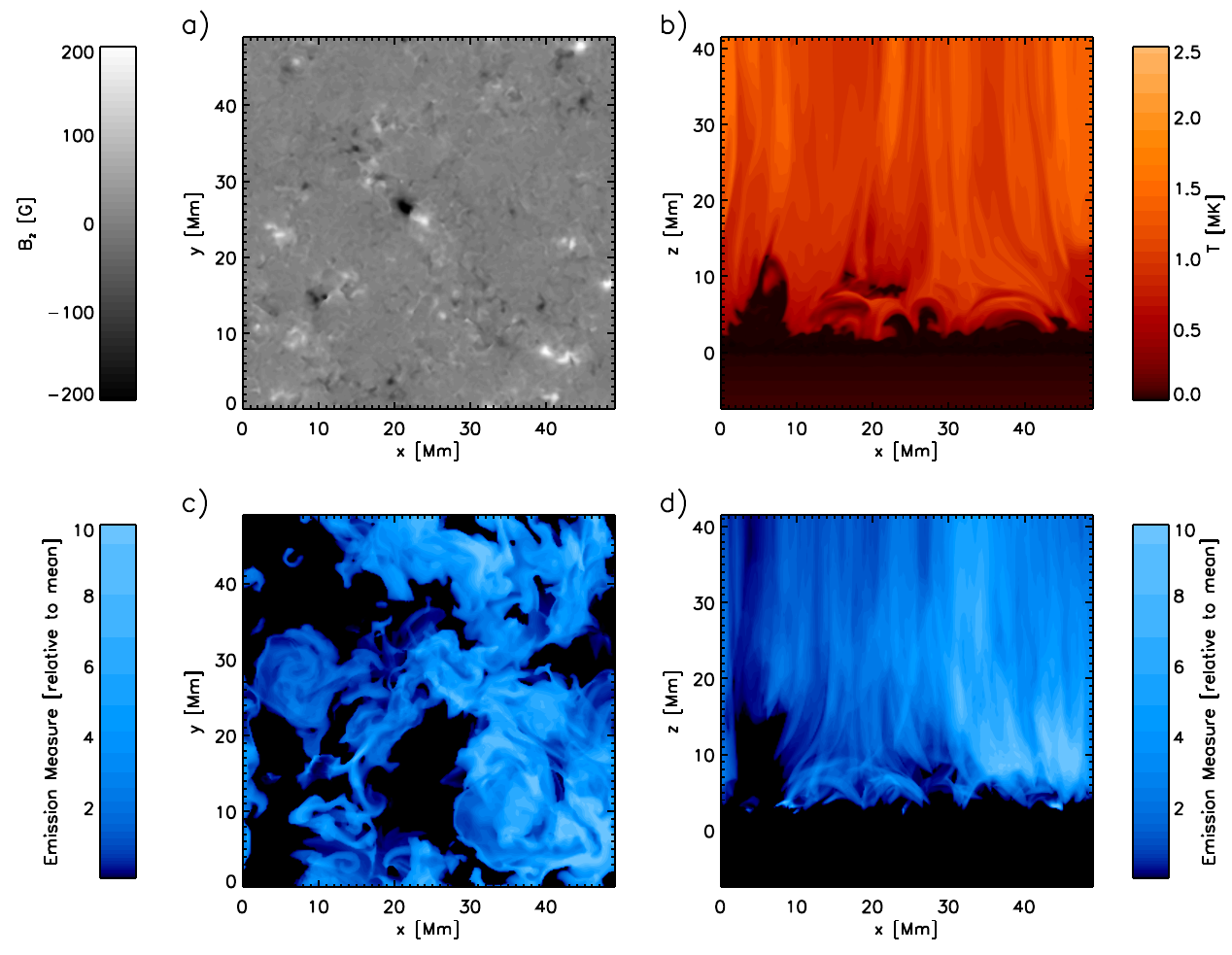}}
   	\caption{Same as Figure \ref{fig:4} for the open flux (OF) setup. This figure is available as an animation.}
   	\label{fig:5}
\end{figure*}

\begin{figure*}
  	\centering
   	\resizebox{0.95\hsize}{!}{\includegraphics{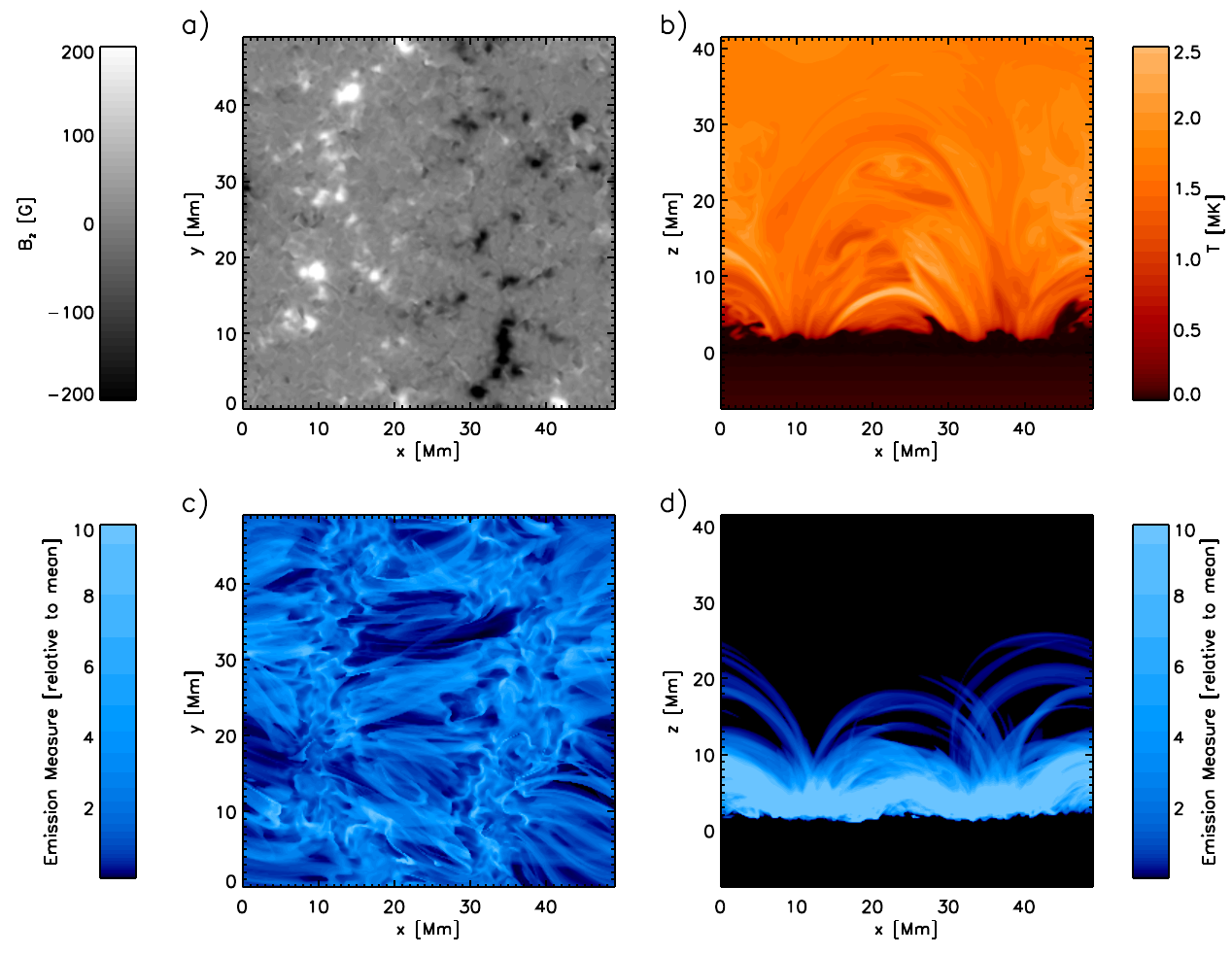}}
   	\caption{Same as Figure \ref{fig:4} for the coronal arcade (CA) setup. This figure is available as an animation.}
   	\label{fig:6}
\end{figure*}

\begin{figure*}
  	\centering
   	\resizebox{0.95\hsize}{!}{\includegraphics{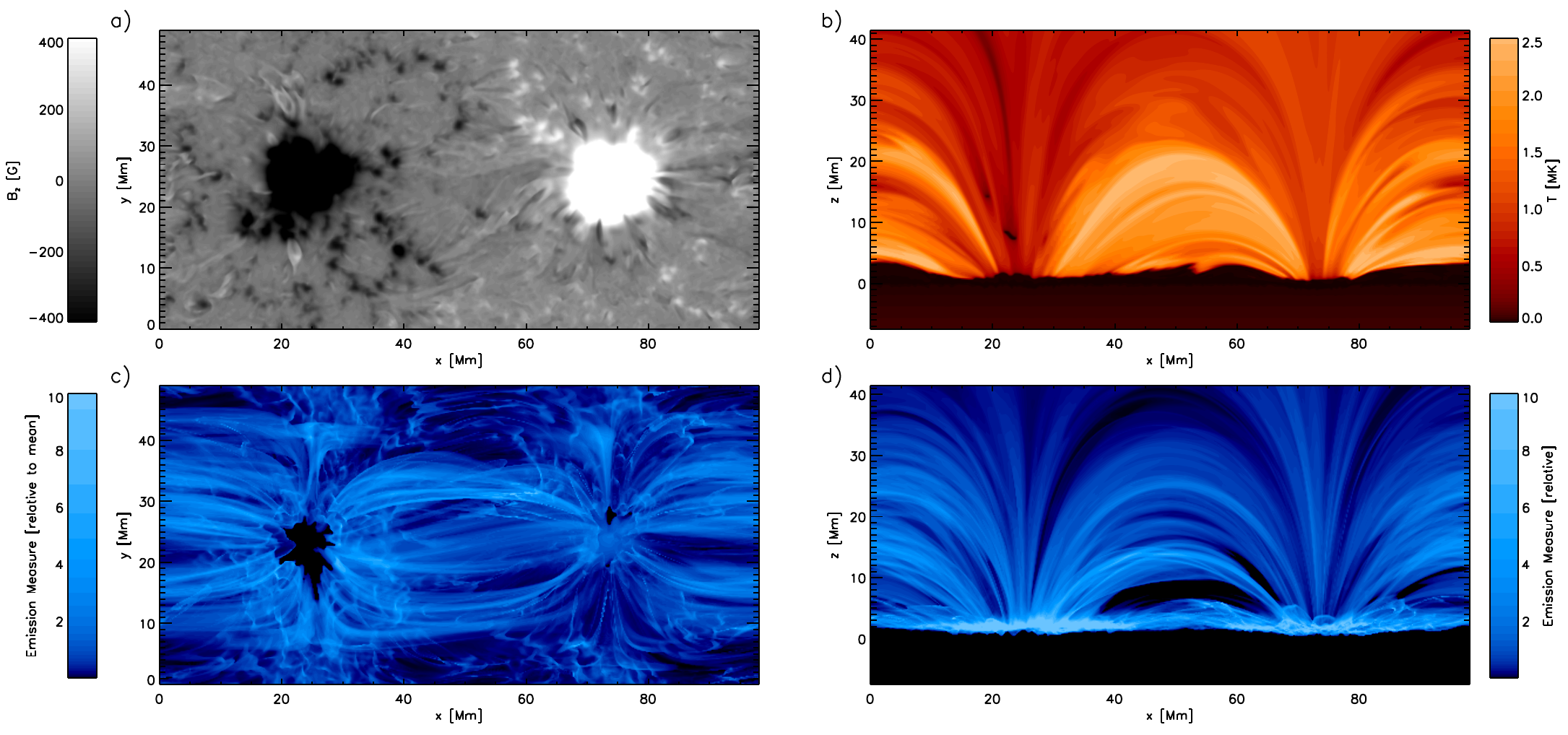}}
   	\caption{Same as Figure \ref{fig:4} for the active region (AR) setup. This figure is available as an animation.}
   	\label{fig:7}
\end{figure*}

\begin{figure}
 	\centering   	
 	\resizebox{0.66\hsize}{!}{\includegraphics{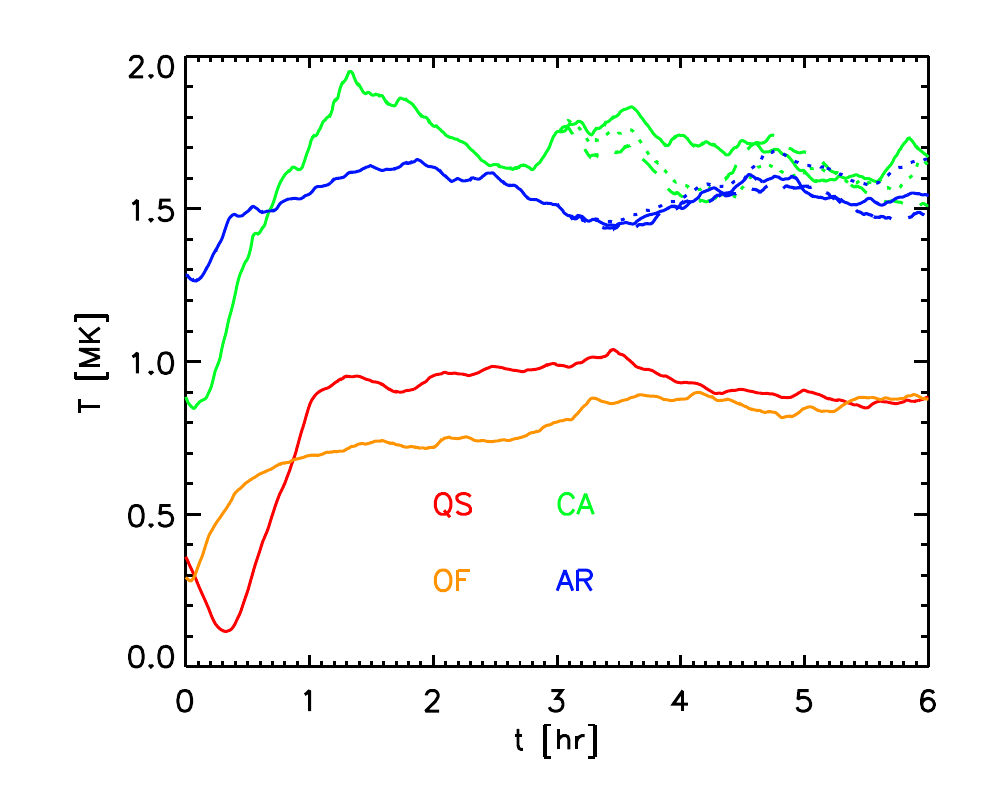}}
   	\caption{Time evolution of the coronal mean temperature averaged in regions with densities $\varrho < 10^{-12}$~g~cm$^{-3}$. The line color differentiates the QS (red),
   		OF (orange), CA (green) and AR (blue) setups. For the CA setup we computed in addition to our $P_{\rm m}>1$ reference case control experiments with 
   		$P_{\rm m}\sim 1$ (green, dotted) and $P_{\rm m}<1$ (greed, dashed). For the AR setup the dotted/dashed lines indicate control 
   		experiments that were computed with a different maximum Alfv{\`e}n velocity: $c=400$~km~s$^{-1}$ (blue, dotted), $c=1600$~km~s$^{-1}$ (blue, dashed).}
  	\label{fig:8}
\end{figure}

\subsection{From quiet to active Sun}
For validating our approach we use four different coronal setups ranging from a quiet Sun to an active region corona. For all simulations we use a domain ranging from about $8$~Mm
beneath the photoshere to about $41$~Mm above the photosphere, i.e. a total vertical extent of $49.152$~Mm. We consider a $49.152^3$ Mm$^3$ domain with $192\times 192\times 64$ km$^3$ grid spacing
for our quiet Sun, open flux and coronal arcade setup. The quiet Sun setup has vertical zero netflux and only a mixed polarity field from a small scale dynamo, the open flux setup has in addition a $3$ G vertical mean
field (i.e. a magnetic net flux of about $7.25\times 10^{19}$~Mx), while we added in the coronal arcade setup a $50$ G vertical mean field in the left and $-50$ G mean field in right half of the domain in the horizontal x-direction (the corresponding large scale flux imbalance is $\pm 6\times 10^{20}$~Mx).

For the active region setup we consider a domain of the size $98.384\times 49.152\times 49.152$ Mm$^3$. Here we started again from our quiet Sun snapshots, but added a pair of opposite polarity
Sunspots, each with a flux of $3.4\times 10^{21}$~Mx and a moderate asymmetry in terms of field strength and coherence, with the spot on the right side being the more coherent one.

In the following discussion we refer to these cases as QS (quiet Sun), OF (open flux), CA (coronal arcade) and AR (active region).

All four setups were first evolved for a few hours without corona with a top boundary just 700~km above the photosphere in order to reach a relaxed state of photospheric magnetoconvection.
In an intermediate step we moved the top boundary 8~Mm above the photosphere using a potential field extrapolation for the magnetic field and a hydrostatic isothermal atmosphere starting with the mean 
pressure and temperature found at the top boundary of the previous run. We use this intermediate step to minimize transients that would occur if we impose a hot corona right on top of our simulation domain
reaching only $700$~km above the photosphere. After
the formation of a transition region reaching at least a few 100,000 K we expanded the domain in a final step to the full extent reported above. While the QS and OF setups reach a statistically steady state, 
the CA and AR setups do show a slow decay. 

Figures \ref{fig:4} (QS), \ref{fig:5} (OF), \ref{fig:6} (CA), and \ref{fig:7} (AR) show snapshots from the four coronal setups described above. We present (a) vertical magnetic field $700$~km above $\tau=1$,
(b) temperature on a vertical cut through the center of the domain, (c) emission measure (line of sight integrated $\varrho^2$) in the $z$-direction, and (d) emission measure in the $y$-direction. We present
the emission measure in the $\mbox{lg}(T/K)=6.0-6.1$ range for all cases. The emission measure is shown on a logarithmic scale with a dynamic range of $1000$ ranging from $0.01$ to $10$ times the mean values
found for the vertical views in each case. The QS and OF cases are barely distinguishable in the magnetogram, since the imposed $3$~G mean vertical field in the
latter is small compared to the shown range of $\pm 200$~G. In the case of the CA setup the positive/negative polarity preference in the left/right half is clearly visible. The temperature and emission
measure show in the QS case a structure of disorganized low lying loops, whereas the OF case is dominated by vertically aligned structures in more than $10$~Mm height. The CA case shows a more
organized structure of loops connecting the opposite polarities. On average the CA case is with $2$~MK about 2 times hotter than the QS and OF cases ($1$~MK).
In the AR case the temperature and emission measure show a clear indication of an organized loop structure connecting both spots. While the average temperature of the
corona is lower than the CA setup, we find higher peak values of more than $5$~MK. For the QS and OF setup the total radiative loss from the corona is about $5\times 10^5$ erg~cm$^{-2}$~s$^{-1}$, for the
CA setup we find $2\times 10^6$ erg~cm$^{-2}$~s$^{-1}$, and for the AR setup $4\times 10^6$ erg~cm$^{-2}$~s$^{-1}$. These values are within the range of $3\times 10^5$ to $10^7$~erg~cm$^{-2}$~s$^{-1}$
for the total energy loss from quiet to active Sun coronae as reported by \citet{Withbroe:Noyes:1977:corona_energy}.

The time evolution of the average corona temperature is shown in Figure \ref{fig:8} (we present here the average in regions with a mass density of less than $10^{-12}$~g~cm$^{-3}$). Starting from 
our initial state the solutions require about $1.5$ hours before they settle into a close to stationary state. In the following comparison we consider for all cases averages from $t=4$ to $6$ hours. 
We show here the QS (red), OF (orange), CA (green) and AR(blue) case.

One key feature of our approach is the artificial limitation of Alfv{\'e}n velocity. For the solutions discussed in the following we use the ``dynamic'' setting of $c=\mbox{max}(C_S,3\,v)$, but also imposed a 
minimum value of $c=400$~km~s$^{-1}$ for the QS, OF and CA cases and a minimum value of $c=800$ ~km~s$^{-1}$ for the AR case. Whereas the AR case stayed close to the imposed $c=800$ ~km~s$^{-1}$ 
minimum, the QS, OF and CA cases showed a few dynamic phases reaching values of $c$ in the $c=400-800$ ~km~s$^{-1}$ range. For the AR case we present additional control experiments using a fixed
value of  $c=400$ ~km~s$^{-1}$ (blue dotted) and  $c=1600$ ~km~s$^{-1}$ (blue dashed), which are further discussed in Section \ref{sect:va}.

For the CA setup we also present two control experiments in which we change the numerical diffusivity in order to study the effect of numerical dissipation on the net coronal heating in our model. We consider
here two setups that have different numerical magnetic Prandtl numbers, $P_{\rm m}$. In addition to our baseline case with $P_{\rm m}>1$ we present a $P_{\rm m}\sim 1$ (green dotted) and $P_{\rm m}<1$ (green dashed) 
case. We further analyze these cases in Section \ref{sec:pm} and quantify the magnitude of numerical diffusivities in Section \ref{sec:est_num_diff}.

\begin{figure*}
  	\centering
   	\resizebox{0.95\hsize}{!}{\includegraphics{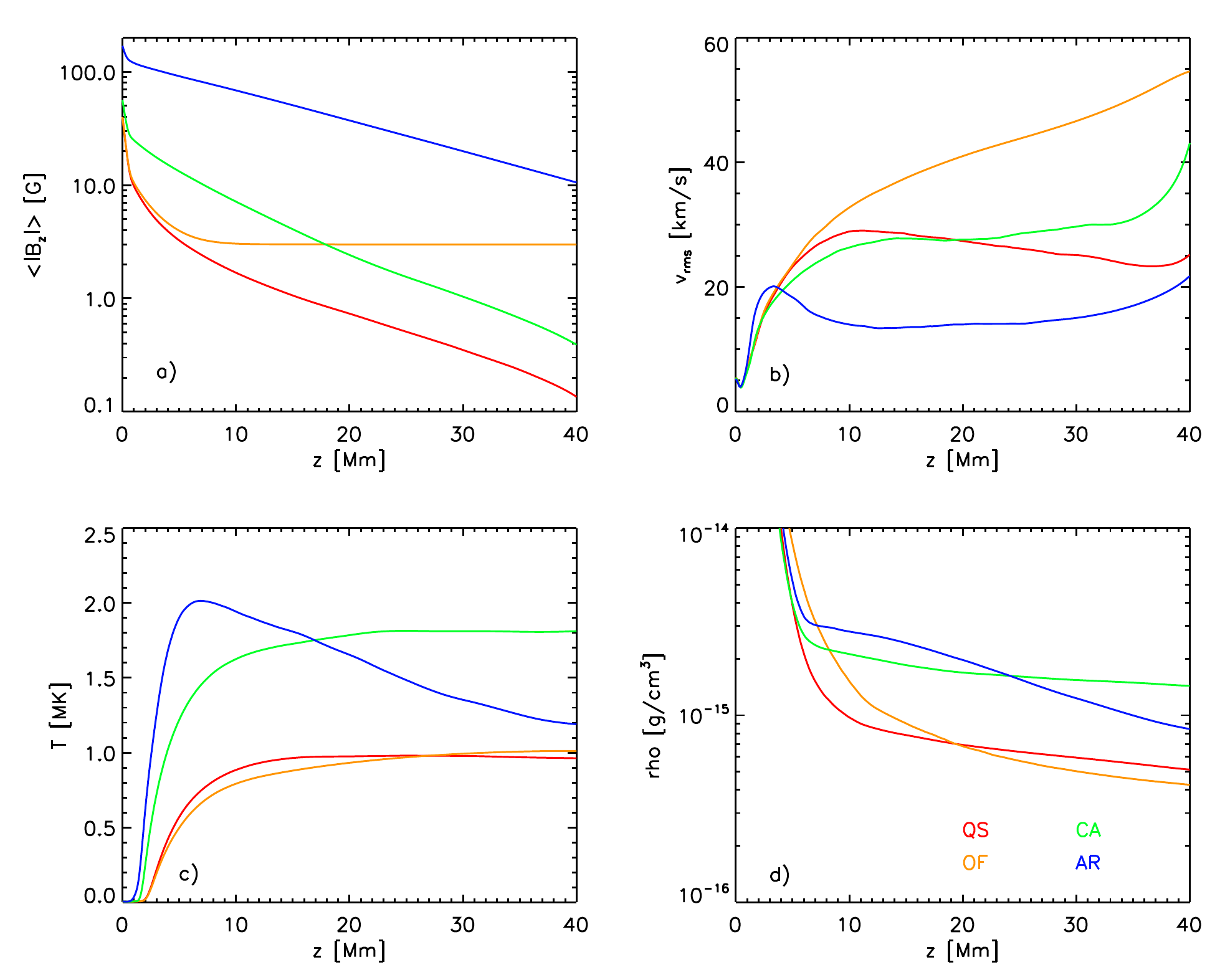}}
   	\caption{Comparison of horizontally averaged quantities for the QS (red), OF (orange), CA (green) and AR (blue) solutions. 
   		a) mean vertical magnetic field strength, b) RMS velocity, c) Temperature and d) density as function of height. $z=0$ Mm corresponds to the position of
   		the photosphere.}
   	\label{fig:9}
\end{figure*}

\begin{figure*}
  	\centering
   	\resizebox{0.95\hsize}{!}{\includegraphics{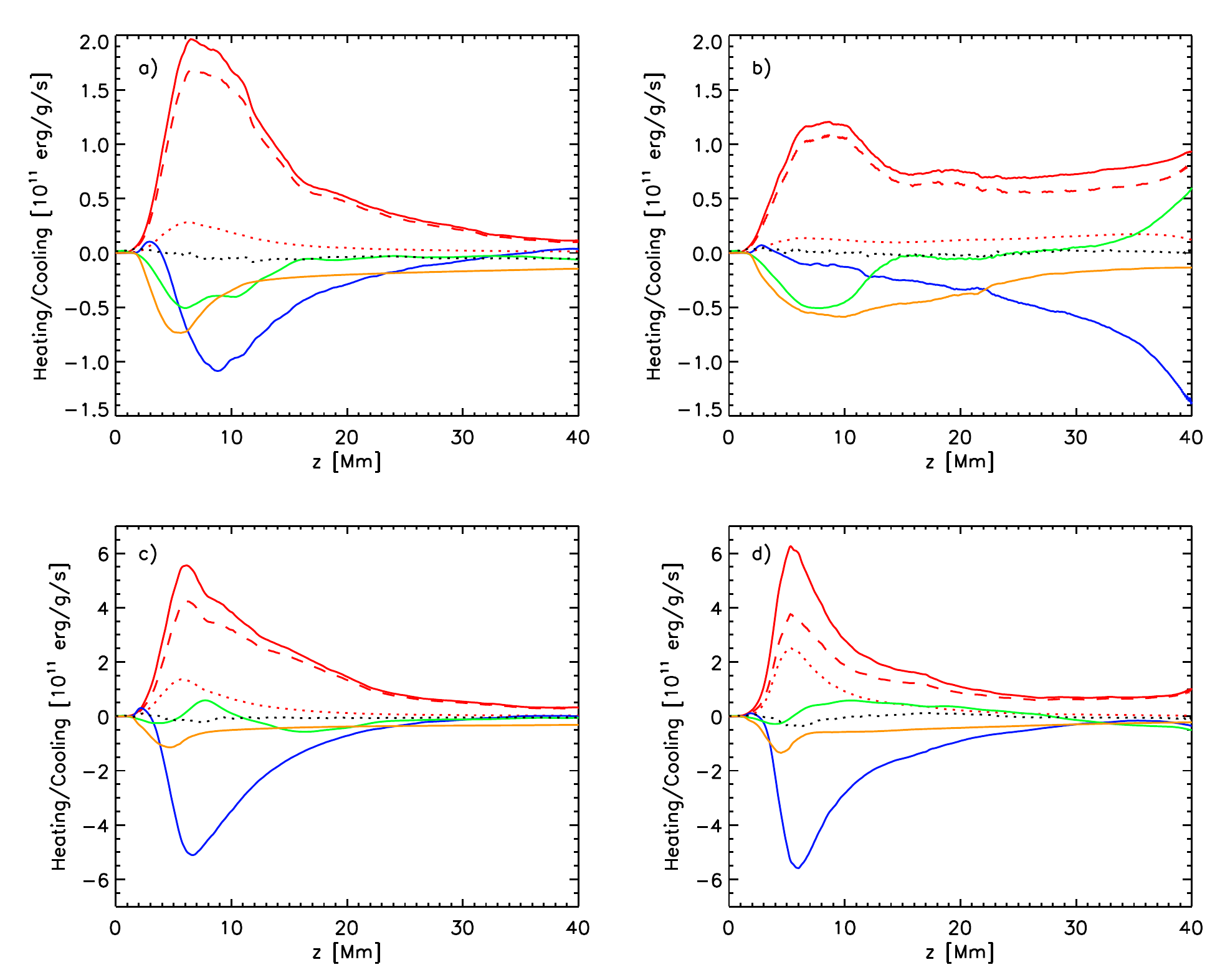}}
   	\caption{Comparison of the energy balance for the the QS (a), OF (b), CA (c) and AR (d) solutions. Shown are the total magnetic 
   	energy input (red, solid), the contribution from resistive heating alone (red, dotted), the work done by the Lorentz force (red, dashed), heat conduction (blue), 
   		advective energy flux (green) and radiative loss (orange). The black dotted line shows the sum of all contributions.}
   	\label{fig:10}
\end{figure*}

\begin{figure}
  	\centering
   	\resizebox{0.66\hsize}{!}{\includegraphics{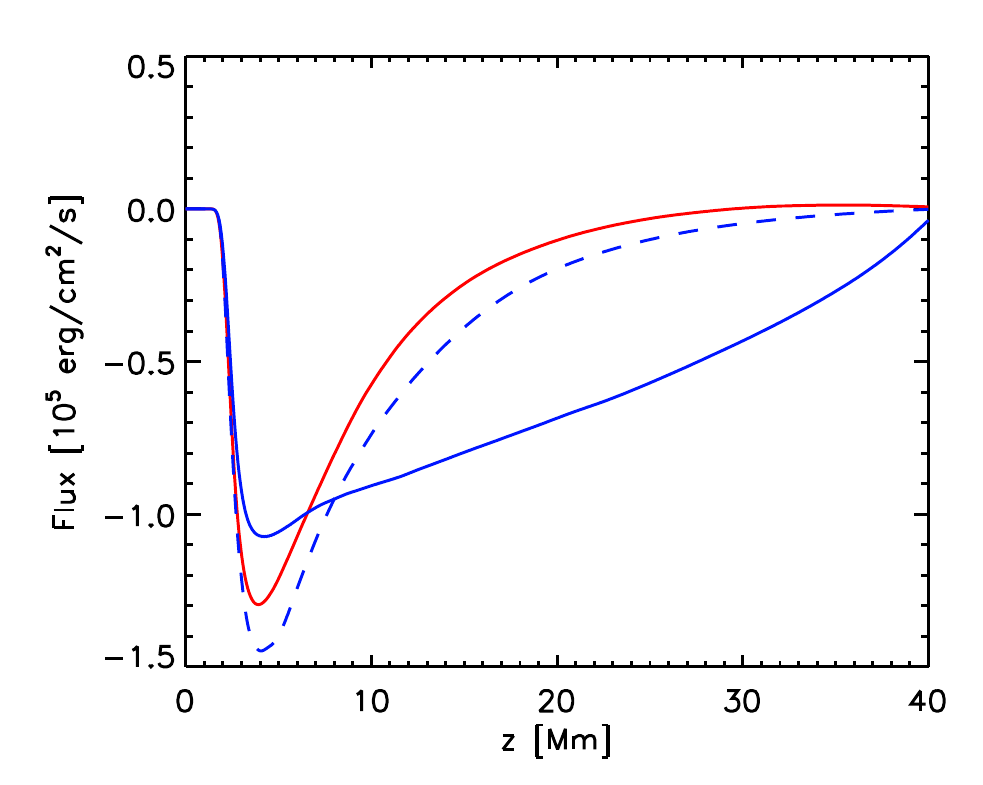}}
   	\caption{Comparison of the conductive heat flux for the quiet Sun (QS, red) and open flux (OF, blue) solution. The blue dotted line shows the heat flux of the
   		open flux case rescaled by $(T_{\rm QS}/T_{\rm OF})^{3.5} \langle |B_z|\rangle_{\rm QS}/\langle |B_z|\rangle_{\rm OF}$, which accounts for most of the differences
   		in the upper part of the simulation domain.}
   	\label{fig:11}
\end{figure}

\begin{figure*}
  	\centering
   	\resizebox{0.95\hsize}{!}{\includegraphics{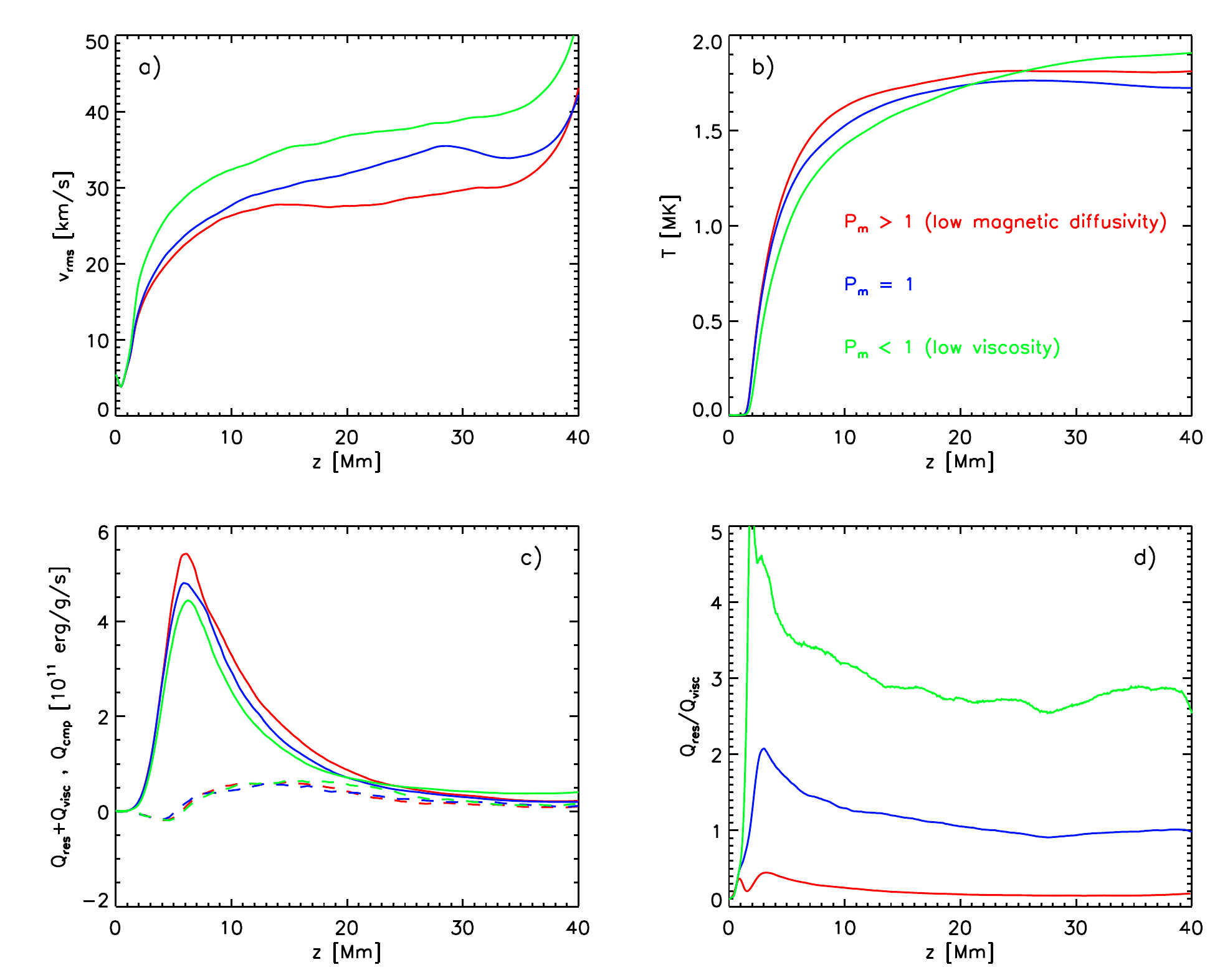}}
   	\caption{Dependence of the coronal mean structure on the magnetic Prandtl number for the coronal arcade (CA) setup:  $P_{\rm m}>1$ due to low magnetic diffusivity (red), 
   		$P_{\rm m}\sim 1$ (blue) and $P_{\rm m}<1$ due to low viscosity (green). We compare the following quantities: (a) RMS velocity, (b)  mean temperature 
   		(c) sum of viscous and resistive heating(solid) and compressional heating (dashed), and (d) the ratio of resistive and viscous heating.
   		$P_{\rm m}$ affects mostly the ratio of resistive and viscous heating, while the sum of both remains similar. As a consequence the influence
   		on the coronal temperature is moderate.}
   	\label{fig:12}
\end{figure*}

\subsection{Coronal energy balance}
In Figure \ref{fig:9} we compare the horizontally and temporally averaged profiles of mean vertical field strength (a), RMS velocity (b), Temperature (c), and density (d). The QS, OF and CA solutions
have a similar mean vertical field strength in the photosphere, but differ significantly in the coronal part of the simulation domain. The QS and OF solutions are similar up to a height of about $8$~Mm
after which the imposed $3$~G  mean field dominates. The CA solution dominates over the OF solution in terms of mean vertical field strength to $20$~Mm. Qualitatively the QS, OF and CA setups have very similar coronal mean temperature profiles, but
differ significantly in the temperatures reached. The AR setup has the hottest lower corona, but has a cooler upper corona compared to the CA setup. This is due to the confinement of hot loops to lower
heights \citep{Chen:etal:2015NatPh,Cheung:etal:2015:DEM}, where they connect to regions with enhanced Poynting flux at the periphery of the spots. The upper part of the simulation is mostly magnetically connected
to the spot umbrae with relatively low Poynting flux. With exception
of the very low corona the AR setup has also the lowest RMS velocities of the four setups considered. However, the highest peak temperatures and peak flow velocities are found in the AR setup. 

Figure \ref{fig:10} presents for the four setups the balance of the different terms in the energy equation. Here we consider the terms that appear in the plasma energy (i.e. internal + kinetic) equation Eq. \ref{eq:ener},
which we solve numerically. Solid red lines show the total magnetic energy input, consisting of resistive heating (red, dotted) and Lorentz force work (red, dashed). Blue lines show the divergence of the conductive
energy flux, green lines the divergence of the advective (i.e. enthalpy + kinetic) energy flux. Radiative losses are shown by the orange line. The black dotted line indicates the sum of all terms. All terms are shown
per unit mass, i.e. we show their horizontal average divided by the horizontally averaged mean density. Panels (a) to (d) show the QS, OF, CA, and AR setups, respectively. A common feature
in all 4 setups is the peak of the magnetic heating in the lower corona around a height of $5-10$~Mm. In all cases the energy input is dominated by Lorentz force work (red, dashed), which is a consequence
of choosing a high numerical magnetic Prandtl number setup. We will discuss the role of $P_{\rm m}$ further in Section \ref{sec:pm}. 
The CA and AR setups reach peak heating rates about three times as large as the QS setup, which is reflected in the overall higher mean temperature.
Compared to the CA setup, the heating in the AR setup is more concentrated towards the lower corona. This explains the steeper rise of the mean temperature in the transition region and the lower 
temperature in the upper parts of the simulation domain, where it is not very different from the QS setup. The QS case shows a magnetic energy input profile that is very similar to the
CA setup apart from the difference in amplitude by a factor of three. 

The OF setup shows a significantly larger total magnetic energy input than the QS setup above $20$~Mm height. However, this does not translate
into a larger mean temperature, since at the same time cooling by conduction is far more efficient in the OF case. While conductive cooling peaks in all other cases in the lower corona, it increases towards the top
boundary in the OF setup. This different behavior is explained by the close to constant unsigned magnetic flux in the OF setup. Here heat conduction is very efficient and operating mostly in the vertical direction.
In all other cases the heat flux is channeled along loops and those field lines that connect to the upper parts of the simulation domain channel the heat flux through footpoint with a small filling factor in the lower corona.
This effect is illustrated in Figure \ref{fig:11}, where we compare the vertical conductive heat fluxes in the QS (red line) and OF (blue line) setups. Even though both cases have a similar mean temperature profile, the
conductive heat flux in the OF setup is larger than the QS case. Rescaling the conductive heat flux in the OF case by $(T_{\rm QS}/T_{\rm OF})^{3.5} \langle |B_z|\rangle_{\rm QS}/\langle |B_z|\rangle_{\rm OF}$
(dashed blue line) accounts for most of the differences in the upper part of the simulation domain. We note that our mostly closed top boundary is not the most appropriate choice for an open flux region and 
may result in an overestimation of heating. According to \citet{Withbroe:Noyes:1977:corona_energy} most
of the energy input in the OF setup should lead to acceleration of solar wind, which is clearly beyond the scope of our simulations due to the rather restricted domain extent with height.

\subsection{Dependence on numerical diffusivity and magnetic Prandtl number}
\label{sec:pm}
Since we do only use numerical diffusivities in our code we study here how they influence the energy dissipation in the corona and the resulting net coronal heating. Using a pure MHD system we 
have two dissipative processes: resistive and viscous dissipation. We computed two additional control experiments for the CA setup in which we changed the effective numerical magnetic Prandtl number 
by combining different numerical diffusivities for the momentum and induction equation. While we used in our baseline case a combination of $h=1.25 [5]$ for $v [B]$, leading to a high numerical $P_{\rm m}$, 
we compute a moderate $P_{\rm m}$ setting with $h=1.25 [1.25]$ and a low $P_{\rm m}$ setting with $h=5 [1.25]$. We apply these changes only in regions with $\varrho<10^{-11}$~g~cm$^{-3}$, i.e. the convection
zone part of our solution remains unaffected. We note that even using the same numerical dissipation scheme for velocity and magnetic field does not necessarily imply the same effective diffusivity. For the corona
the numerical scheme is in general less diffusive for the magnetic field compared to the velocity field, since the low $\beta$ condition leads to a rather smooth magnetic field regardless of the value of diffusivity, i.e. the solution is biased towards
a higher numerical  $P_{\rm m}$. Estimating effective diffusivities as described in Section \ref{sec:est_num_diff} leads to numerical $P_{\rm m}$ values of about $49$, $3.7$ and $0.27$ for the three cases discussed above.

It was found by \citep{Brandenburg:2011:SSD_low_Pm,Brandenburg:2014:Pm} that the magnetic Prandtl number determines the partitioning between resistive and viscous dissipation in MHD turbulence and dynamos
(high-$\beta$ regime). In the low $P_{\rm m}$ solar convection zone most of the dissipation is expected to happen through resistivity, whereas the opposite is the case for the high $P_{\rm m}$ regime. We study here 
the dependence of (numerical) coronal heating on the (numerical) magnetic Prandtl number in order to evaluate if comparable trends are also found in the low-$\beta$ regime of the solar corona 

The time evolution of these control experiments is shown in Figure \ref{fig:8}, in Figure \ref{fig:12} we compare time averages from $4-6$ hours
 in terms of  RMS velocity (panel (a)), mean temperature (panel (b)), total heating (sum of viscous and resistive heating as well as compressional heat) (panel (c)) and the ratio of
resistive to viscous heating (panel (d)). Except for the latter, which varies dramatically with $P_{\rm m}$, the other quantities show only a moderate variation. In view of the intrinsic variation of the coronal mean temperature
with time as shown in Figure \ref{fig:8}, we would require a substantially longer time series in order to determines if the difference shown in panel (b) are actually statistically significant.  
 
Overall we find that, similar to the high-$\beta$ convection zone \citep{Brandenburg:2011:SSD_low_Pm,Brandenburg:2014:Pm}, the magnetic Prandtl number also determines the partition between resistive and viscous
heating in the low-$\beta $ corona. This result is non-trivial, since the underlying energy fluxes through the system are quite different. In the high-$\beta $ convection zone pressure/buoyancy driving provides the
primary energy input into the system. A fraction of that energy is directly dissipated through viscosity, while the rest of it is converted through the Lorentz force into magnetic energy (small- and/or large-scale dynamo) and
subsequently dissipated through resistivity. Here $P_{\rm m}$ modulates the efficiency of the dynamo process. In the low-$\beta $ corona the system is primarily driven by the buildup and release of magnetic stresses.  Here
$P_{\rm m}$ modulates the partition of the energy release in terms of resistive heating and Lorentz force work.

Based on the $P_{\rm m}$ dependence we find, direct resistive heating is expected to be insignificant for coronal conditions ($P_{\rm m}\sim 10^{10}$) and almost all energy release should
happen through the Lorentz force driving flows, which eventually thermalize in shocks \citep[see, e.g.][]{Longcope:etal:2009,Guidoni:Longcope:2010}, or just due to viscous stress. Running a simulation in a high
$P_{\rm m}$ regime (or at least at the highest value numerically feasible) allows to capture this two stage process, although the details of post reconnection flows and shocks cannot be captured with our current resolution.
Most importantly, $P_{\rm m}$ changes only the partition between resistive and viscous heating while the sum remains mostly unchanged. As a consequence the resulting coronal mean temperatures are mostly $P_{\rm m}$
(i.e. dissipation process) independent. While resistive and viscous heating do not necessarily show the same spatial distribution, these differences are effectively smoothed by heat conduction.  

\subsection{On the magnitude of numerical diffusivities}
\label{sec:est_num_diff}
We estimate the amplitude of numerical diffusion terms for the CA setup. Quantifying an effective numerical diffusivity is in general non-trivial, since the numerical resistive and viscous heating do not necessarily have a 
functional form that can be easily compared to the expressions of explicit resistivity and viscosity given by:
\begin{eqnarray}
	\epsilon_{\nu}&=&\nu\varrho \sum_{i,k}\frac{\partial v_i}{\partial x_k}\left[\frac{\partial v_i}{\partial x_k}+\frac{\partial v_k}{\partial x_i}-\frac{2}{3}\delta_{ik}\nabla\cdot\vec{v}\right]\\
	\epsilon_{\eta}&=&\frac{\eta}{4\pi} \vert\nabla\times\vec{B}\vert^2
\end{eqnarray}

\begin{deluxetable}{  c  c  c  c }
  \tablecaption{Effective diffusivities and effective magnetic Prandtl numbers for the high, moderate and low $P_{\rm m}$ CA setups.}
  \tablehead{
  \colhead{Setting} & \colhead{$\eta_{\rm eff} [$~cm$^2$s$^{-1}]$} &  \colhead{$\nu_{\rm eff}[$~cm$^2$s$^{-1}]$} & \colhead{$P_{\rm m\, eff}$}
  }
  \startdata
  high $P_{\rm m}$          & $2.9\times 10^{11}$      &  $1.4\times 10^{13}$  & 49 \\
  moderate $P_{\rm m}$ &  $2\times 10^{12}$  &   $7.4\times 10^{12}$  & 3.7 \\
  low $P_{\rm m}$           & $2.4\times 10^{12}$   &  $6.6\times 10^{11}$  & 0.27
  \enddata
  \label{tab:pm}
\end{deluxetable}

For example, we find that for the large $P_{\rm m}$ setup numerical resistive heating has only a very poor correlation of about $0.2$ with $\epsilon_{\eta}$. The correlation is only moderately better for the numerical viscous
heating and $\epsilon_{\nu}$ with a value of about $0.45$ . If we nonetheless use these expressions to define effective numerical diffusivities by equating the total value of $\epsilon_{\nu}$ and $\epsilon_{\eta}$ with
the total value of the respective numerical terms (integrated over the volume of the corona), we find for our three cases the values presented in Table \ref{tab:pm}.

These values are significantly smaller than those that would be required in a simulation using only explicit diffusivities. \citet{Bingert:Peter:2011} used values
comparable to $v_{\rm rms}\Delta x$, which were in their setup $\eta=10^{14}$~cm$^2$~s$^{-1}$ and $\nu=10^{15}$~cm$^2$~s$^{-1}$. While our code also uses diffusivities as large as $10^{15}$~cm$^2$~s$^{-1}$, they are 
restricted to regions where they are required for numerical stability. Overall our simulations combined with those presented by \citet{Bingert:Peter:2011,Chen:etal:2014} cover a range of more than two order of magnitude in
diffusivities, which indicates a significant robustness with regard to the treatment of diffusivities in terms of magnitude and functional form (explicit vs. numerical). The value $\nu_{\rm eff}$ we find in our simulation is comparable 
to or even smaller than typical values of the Spitzer 
viscosity under coronal conditions. Eq. (\ref{eq:nu_spitzer}) yields for $T=1.5$~MK, $\varrho=2\times 10^{-15}$~g~cm$^{-3}$ and $\ln\Lambda=20$ a value of  $1.4\times 10^{14}$~cm$^2$~s$^{-1}$. The value of $\eta_{\rm eff}$ is of
course still much larger than the Spitzer value of $5600$~cm$^2$~s$^{-1}$ for those values. Nonetheless, these values indicate in combination with the $P_{\rm m}$ dependence we discussed above that current numerical simulations can 
in principle capture the dominant scale of energy dissipation, provided they can be run in a high enough  $P_{\rm m}$  regime: Under these conditions resistive heating is insignificant and the detailed value of $\eta_{\rm eff}$ does not matter 
as long as it allows for fast reconnection; energy dissipation happens through viscosity on scales that can be captured in current simulations (although the detailed microphysics of this mostly collision less regime are clearly beyond the scope 
of single fluid models).

\begin{figure*}
  	\centering
   	\resizebox{0.95\hsize}{!}{\includegraphics{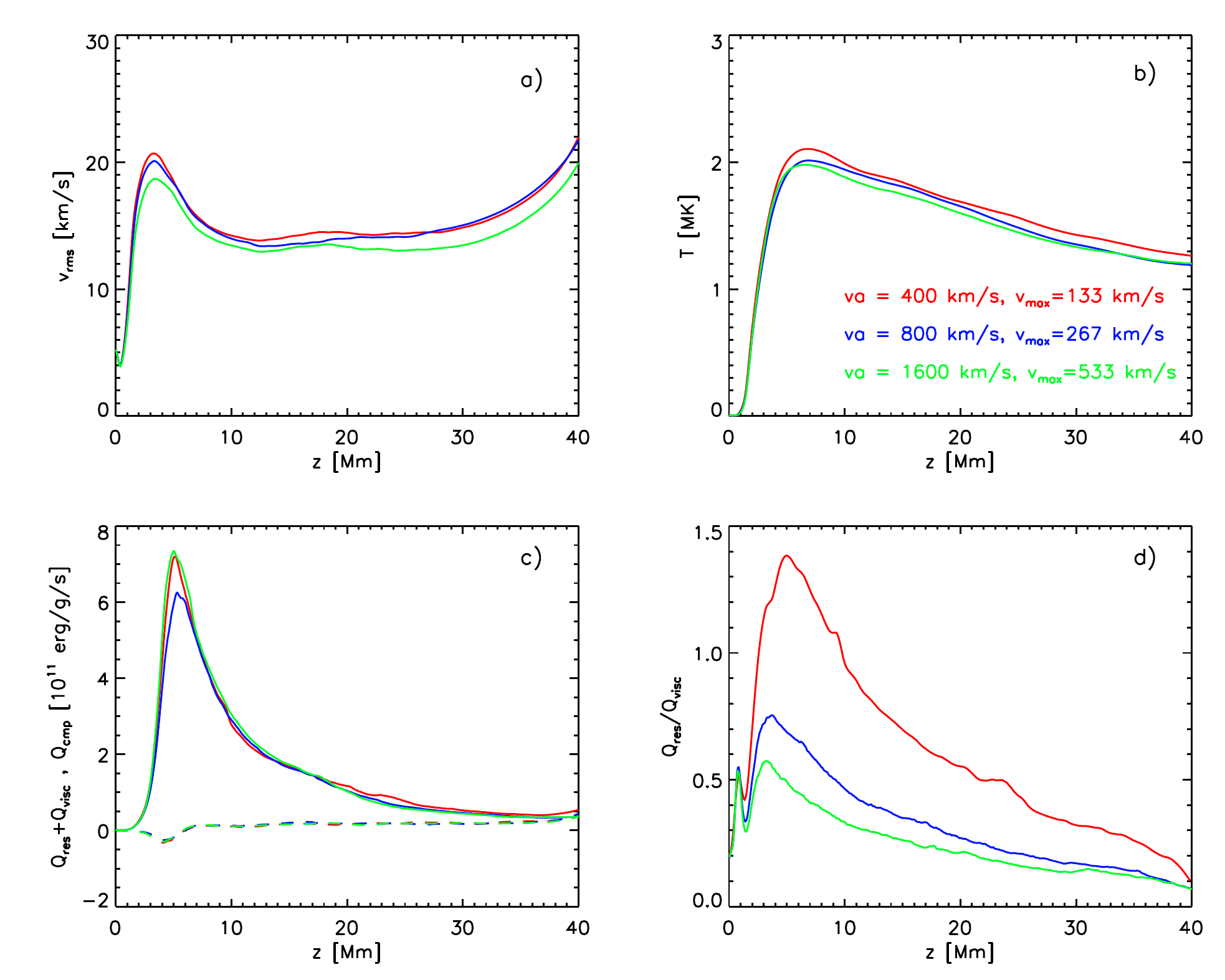}}
   	\caption{Dependence of the AR solutions on the cutoff for the Alfv{\'e}n velocity. We show a solutions with $c=400$ (red), $800$ (blue), and $1600$~km~s$^{-1}$ (green). In all three cases we limit $v$ by $c/3$
   	         in order to ensure the validity of the semi-relativistic approach. Panels a) to d) show RMS velocity, temperature, heating, and ratio of resistive to viscous heating as in Figure \ref{fig:12}. The effect of 
   		$c$ on the coronal energy deposition is very moderate. For low values of $c$ we do see an increase of the ratio of resistive to viscous heating similar to a low $P_{\rm m}$ case.}
   	\label{fig:13}
\end{figure*}

\subsection{Dependence on the maximum Alfv{\'e}n velocity}
\label{sect:va}
We explore the dependence of the solutions on the maximum allowed Alfv{\'e}n velocity (i.e. semi-relativistic treatment) for the active region setup, since there the influence from the relativistic corrections terms is most
significant. The horizontally averaged (unlimited) Alfv{\'e}n velocity does reach in this simulation values around $10,000$ km~s$^{-1}$, while peak values can exceed $100,000$ km~s$^{-1}$. In addition to our reference
solution, which was computed with a value of $c=800$~km~s$^{-1}$, we compute additional control experiments with $c=400$~km~s$^{-1}$ and  $c=1600$~km~s$^{-1}$. We limit in all cases the flow velocity to 
$c/3$ in order to prevent situations where the flow speed would become comparable or even exceed $c$. The time 
evolution of the coronal mean temperature is presented for both experiments in Figure \ref{fig:8}: $c=400$ km~s$^{-1}$ (blue, dotted), $c=1600$ km~s$^{-1}$ (blue, dashed). In Figure \ref{fig:13} we compare horizontally averaged 
quantities similar to Figure \ref{fig:12}, we consider here for all three simulations the time interval from $4$ to $6$ hours (one hour after these cases were restarted from the reference solution at $t=3$~hours).

Similar to the effect of the numerical magnetic Prandtl number we find also here the most significant effect in the ratio of resistive to viscous heating. Lower values of $c$ put a larger constraint on motions perpendicular 
to field lines and limit the work that can be released through the Lorentz force. As a consequence the heating is shifted more towards resistive heating, similar to a low magnetic Prandtl number setting. The effect on the total
(sum of resistive and viscous) heating and consequently on the temperature is small and comparable to the intrinsic variability found in these simulations, i.e. we would require a substantially longer time duration in order to
quantify if any of these differences are statistically significant.

Lower values of $c$ than those considered here would eventually lead to a significant difference, there is however no reason for using values of $c$ lower than the speed of sound from a standpoint of computational speed. 
It is however remarkable that even using values of $c$ comparable to the speed of sound (about $200-400$~km~s$^{-1}$ in this case) does lead to acceptable results. This would not be the case if we would have artificially limited 
the strength of the Lorentz force where $c=C_S$ implies an effective plasma $\beta$ of unity. 

While desirable, it is unfortunately non-trivial to provide here a formal convergence study by computing a reference solution with the correct speed of light. While numerically very expensive that solution would also provide a significantly
more diffusive corona since the numerical diffusivities scale linearly with the maximum characteristic velocity of the system, which would make a direct side-by-side comparison difficult if not meaningless. We explored here values of
$c$ differing by about a factor of $4$, which corresponds to a factor of $16$ in the magnitude of the semi-relativistic correction terms (i.e. inertia perpendicular to field lines) and found only small differences. 

\begin{figure*}
  	\centering
   	\resizebox{\hsize}{!}{\includegraphics{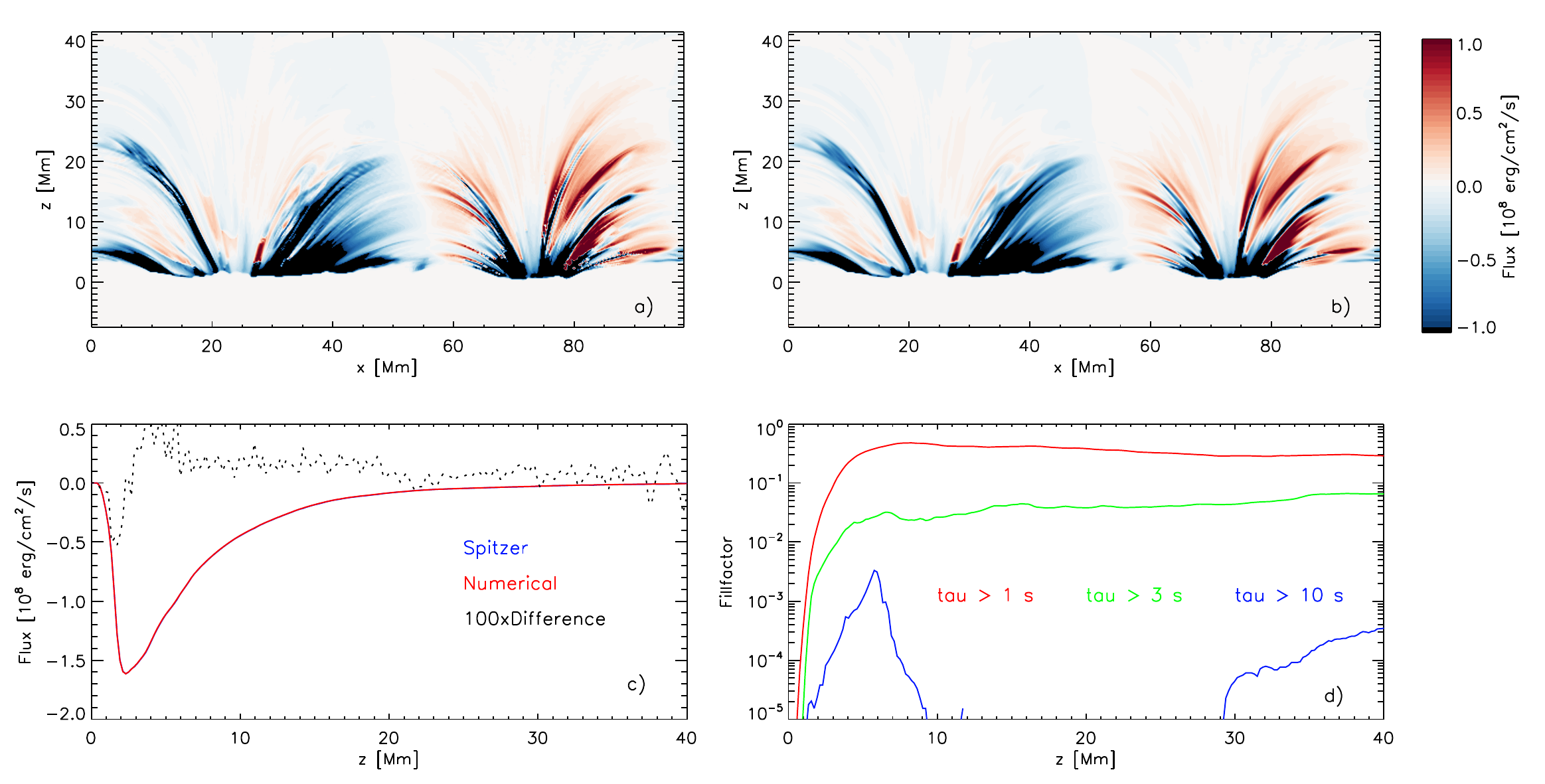}}
   	\caption{Conductive heat flux in the AR setup. Panel a) shows the vertical component of the Spitzer heat flux $-\hat{b}_z  f_{\rm sat}\sigma T^{5\over2}(\hat{\vec{b}}\cdot\nabla)T$ , Panel b) shows the quantity 
   	$q \hat{b}_z$, which we actually compute. Both are presented for a vertical slice through the center of the simulation domain at $y=24.576$~Mm. Panel c) presents the horizontally averaged
   	Spitzer heat flux (blue) and the numerical heat flux (red), the difference of both multiplied by a factor of $100$ is indicated as black dotted line. Panel d) shows the horizontal fill factor of regions with values 
   	of $\tau$ exceeding values of $1$, $3$, and $10$ seconds as function of height.}
   	\label{fig:14}
\end{figure*}

\subsection{Conductive heat flux}
\label{sec:conduction}
In order to validate our treatment of heat conduction we compare for a snapshot from the AR simulation the vertical component of Spitzer heat flux $-\hat{b}_z f_{\rm sat}\sigma T^{5\over2}(\hat{\vec{b}}\cdot\nabla)T$ to
the heat flux $\hat{b}_z q$ we compute according to Eqs. (\ref{eq:ener}) and (\ref{eq:cond}). We use here the same snapshot as highlighted in Figure \ref{fig:7}. For a vertical slice through the simulation domain we present in Figure \ref{fig:14}a) the Spitzer heat flux and in Figure \ref{fig:14}b)
the quantity $\hat{b}_z q$. As expected $\hat{b}_z q$ is smoother than the Spitzer flux since it corresponds formally to a heatflux averaged over a time scale
$\tau$. Nonetheless it captures well the overall structure of the heatflux. Since we compute our simulation with a smoothed heatflux, the corresponding temperature profile is more rough as it would be in a simulation computed with Spitzer conductivity in the first place. As a consequence the Spitzer heat flux computed from a simulation snapshot is more rough as it would be in a simulation that is computed with a Spitzer heatflux directly. Most of the differences do exist on a scale comparable to the grid spacing. Most importantly for the energetic balance, the quantity $\hat{b}_z q$ correctly captures the total conductive heat flux.
Figure \ref{fig:14}c) compares the horizontally averaged vertical heatflux computed from Spitzer, $\langle -\hat{b}_zf_{\rm sat}\sigma T^{5\over2}(\hat{\vec{b}}\cdot\nabla)T\rangle$, (blue) to the quantity $\langle q \hat{b}_z\rangle$
 (red), the difference enhanced by a factor of $100$ is shown as black dotted line. The maximum difference is about $0.3\%$ of the peak heatflux value. Another quantity we can use to characterize how well our treatment of heat conduction works 
 is the averaging timescale $\tau$. Our numerical treatment
 can only capture well features that evolve on a time scale larger than $\tau$. Figure \ref{fig:14}d) presents the horizontal fill factor of regions with values of $\tau$ larger than $1$, $3$, and $10$ seconds as function of height. $\tau$ exceeds a 
 timescale of $10$~seconds in only less than $0.4\%$ of the pixels at any given height, i.e. we do capture well coronal evolution on timescales of interest. The maximum value of $\tau$ in the presented snapshot is $273$~seconds, while our numerical integration timestep is $0.1$ seconds. The diffusive timestep limit is about $2\cdot 10^{-5}$ seconds, i.e. our effective speedup compared to an explicit treatment of heat conduction is about a factor of $5,000$.
 
 \begin{figure*}
  	\centering
   	\resizebox{\hsize}{!}{\includegraphics{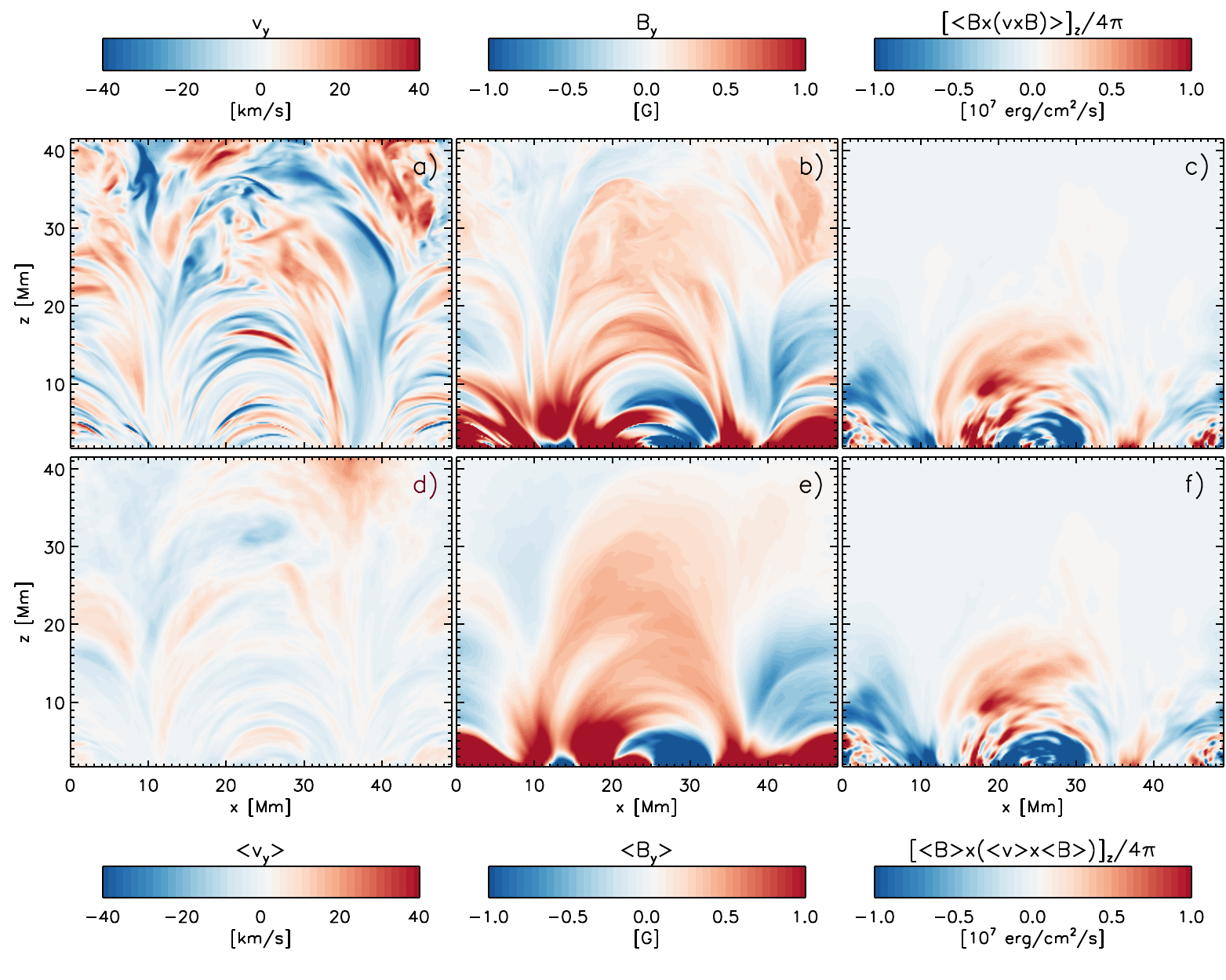}}
   	\caption{Vertical cross section through CA setup showing: a) $v_y$, b) $B_y$ and d) $\langle v_y \rangle$, f) $\langle B_y \rangle$
   		    (components perpendicular to the plane of view), 
   		     where $\langle\ldots\rangle$ denotes a $60$ minute average. Panel c) shows the time averaged vertical component of the     
   		      Poynting flux, while panel f) shows the Poynting flux computed from the time averaged velocity and magnetic field. While
   		       temporal averaging strongly reduces the velocity amplitude, the Poynting flux is less affected, indicating that there is a
   		       significant contribution from motions with long time scales .}
   	\label{fig:15}
\end{figure*}

\begin{figure*}
  	\centering
   	\resizebox{0.75\hsize}{!}{\includegraphics{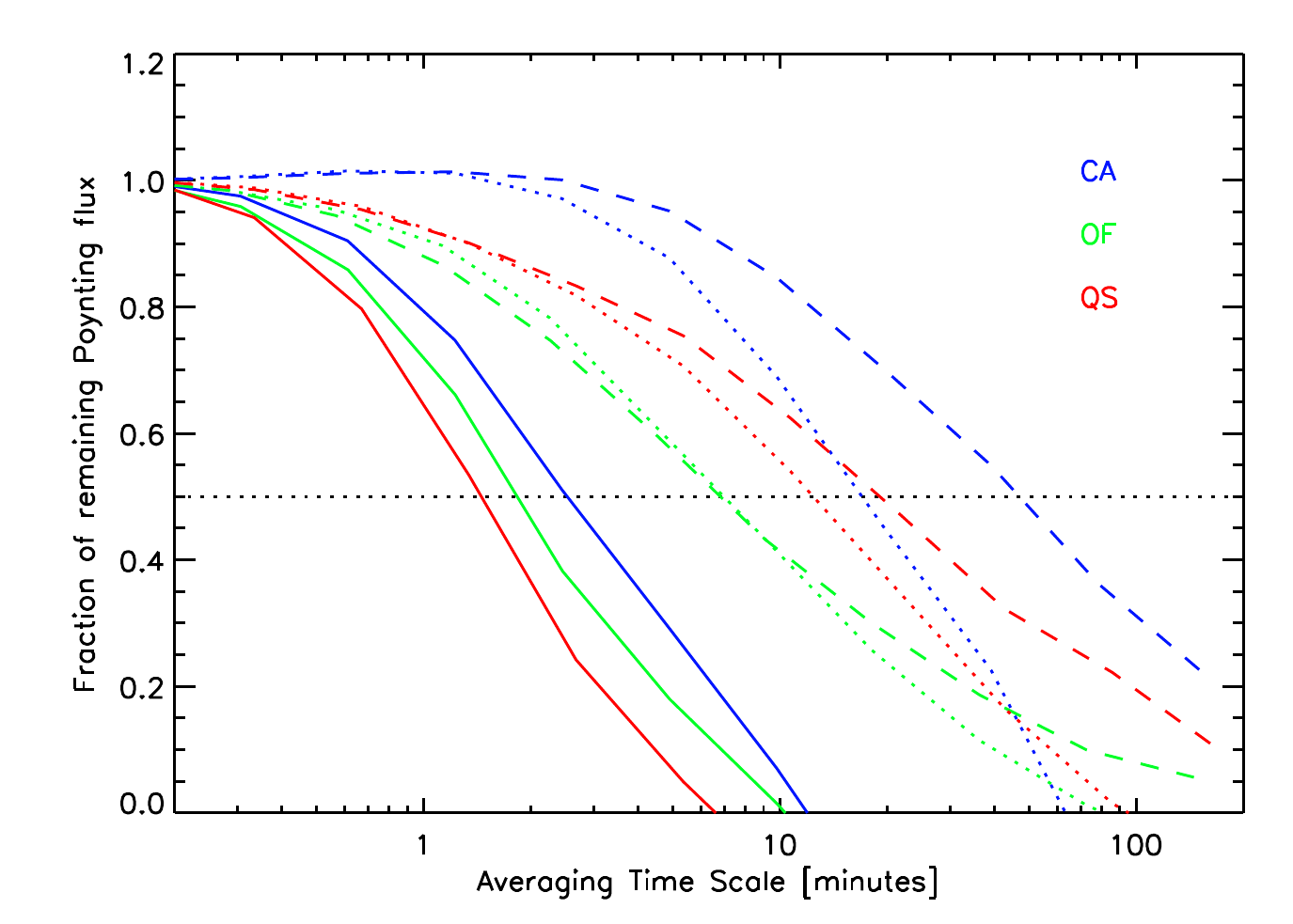}}
   	\caption{Fraction of Poynting flux remaining after application of temporal averaging to velocity and magnetic field. We present
   		results for different constant height surfaces located  $700$~km (solid), $4.8$~Mm (dotted) and $8.9$~Mm (dashed) above the
   		photosphere for the QS (red), OF (green) and CA (blue) setups. Time scales depend on the setup and increase generally 
   		with height. For the CA setup $50\%$ of the Poynting flux in $8.9$~Mm height comes from motions with time scales beyond 50
   		minutes.}
   	\label{fig:16}
\end{figure*}

\begin{figure*}
  	\centering
   	\resizebox{0.75\hsize}{!}{\includegraphics{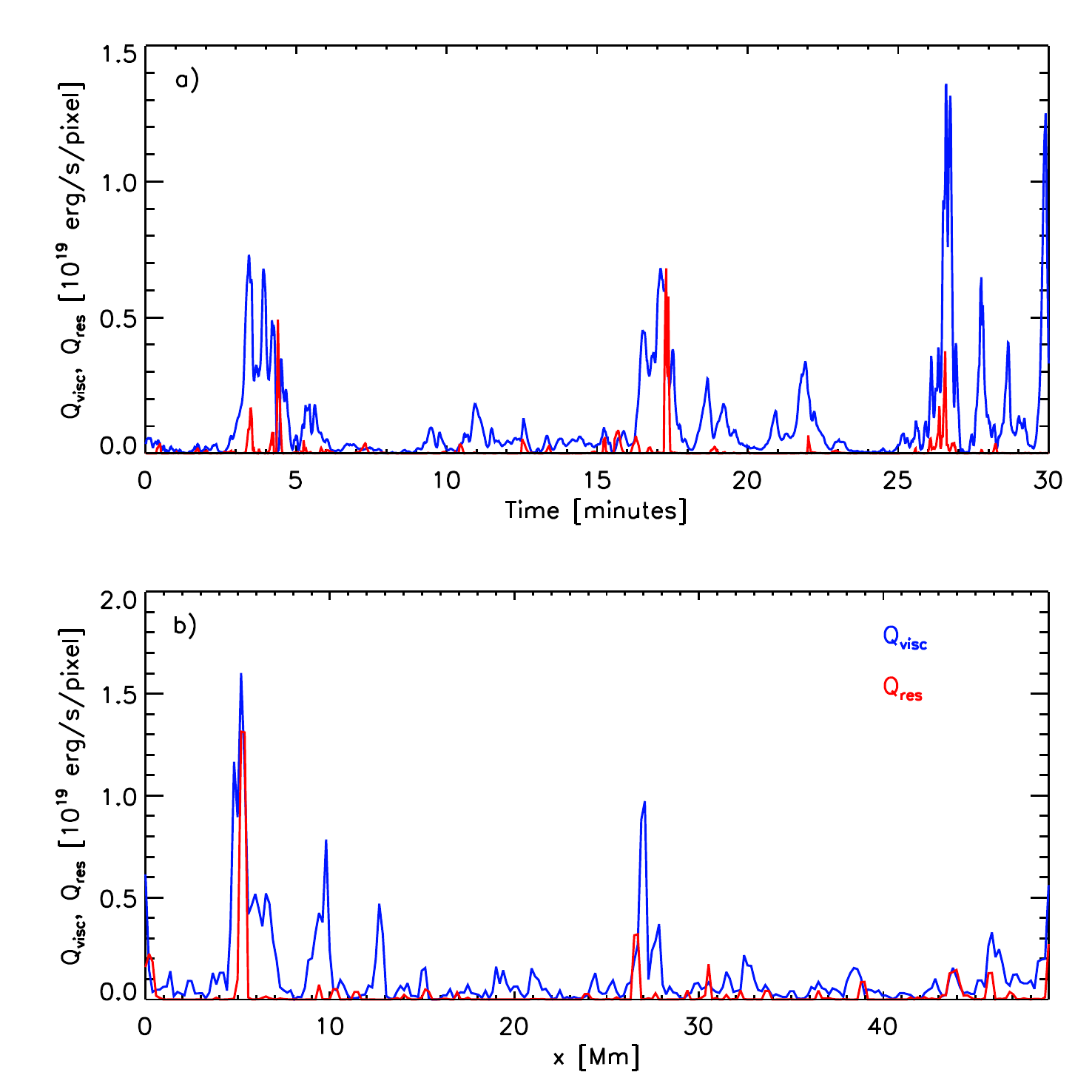}}
   	\caption{Resistive and viscous heating per simulation pixel as function of time for a fixed position in space (a) and as 
   		function of space for a fixed time (b). Resistive heating is shown in red, viscous heating in blue. Both heating terms are 
   		highly intermittent in space and time with typical durations of heating events on the scale of minutes and an extent of a few
   		Mm. The resistive heating shows a higher degree of intermittency since we run the simulation in a high magnetic Prandtl number regime.}
   	\label{fig:17}
\end{figure*}
 
All our test cases concern more or less quiescent corona setups and we found that for our choices of the ``reduced speed of light'', $c$, the hyperbolic treatment of heat conduction appears to be sufficiently accurate. If we would consider setups with significantly higher temperatures, the required
averaging time-scales $\tau$ would increase, while at the same time conductive time scales decrease, i.e. the hyperbolic approach presented here would appear inappropriate for describing conductive heat transport unless we choose much larger values of $c$ (i.e. much smaller time steps). It is however 
important to realize
that Spitzer conductivity itself becomes inappropriate at higher temperatures and can lead to a transport of heat with speeds that are far from physically meaningful. Formally the conductivity is on the order of $\kappa\sim l_e\,v_e$, where $l_e$ is the electron mean free path and $v_e$ the electron thermal velocity.
$v_e$ increases as $T^{1/2}$, while $l_e$ increases as $T^2$. On a grid with spacing $\Delta x$ conduction can lead to effective thermal transport velocities $v_{\kappa}\sim \kappa/\Delta x\sim (l_e/\Delta x)\, v_e$.  In the regime $l_e > \Delta x$ (high temperature) $v_{\kappa}$ can exceed $v_e$, which is physically not meaningful since electrons are responsible for the transport (in addition $v_{\kappa}$ does exceed the (true) speed of light for temperatures above $5$~MK for the grid spacing we use!). Furthermore, considerations of plasma neutrality and the required return currents suggest a saturation flux corresponding to about 
$1/6$ of the electron thermal speed $\sqrt{k_B\,T/m_e}$ \citep[e.g.][]{Fisher:etal:1985:flare,Meyer:2012:super-timestepping}. This saturation electron transport velocity corresponds to about $4\,C_S$ assuming a solar H/He mixture. The effective transport speed is about $2\,C_S$ since only the electrons participate in 
the transport, which leads to a saturation flux as shown in Eq. (\ref{Eq:saturation}). In view of these limitations of the Spitzer conductivity, the hyperbolic treatment appears even appropriate for a higher temperature thermal plasma as long as it allows for transport speeds of about $2\,C_S$ (which can be
easily ensured by choosing the ``reduced speed of light'' in our approach appropriately) and it guarantees by construction that unphysical transport velocities are eliminated from the system. Note that free streaming is by nature more a hyperbolic (advective) than parabolic (diffusive) process. Formally the hyperbolic transport equation (telegraph equation) is a higher order approximation to the Boltzmann equation than the diffusion approximation \citep[see, e.g.][]{Gombosi:1993:telegraph}.

\section{Time scales of coronal energy transport and release}
\label{sec:heating}
The nature of coronal heating has been a subject of heavy debate \citep[see, e.g.][]{Klimchuk:2006:review,Ballegooijen:2014:AC-DC}, most  of the
discussion is focused on the time scales involved in the coronal energy transport and release. Models are typically
classified based on whether the corona responds mostly quasi-statically (``DC''-heating) or very dynamic (``AC'' or wave heating) to photospheric footpoint motions.

We determine the dominant time scales of motions that contribute to the coronal energy transport by comparing the  time averaged Poynting flux 
$\langle \vec{B}\times (\vec{v}\times \vec{B})\rangle/4\pi$ to the Poynting flux computed from temporally averaged
velocity and magnetic field $\langle \vec{B}\rangle \times (\langle \vec{v}\rangle \times \langle \vec{B}\rangle)/4\pi$. While the former
contains contributions from motions of all time scales, the latter lacks the contributions from shorter time scales. We illustrate
the procedure in Figure \ref{fig:15} where we present in panels (a) and (b) $v_y$ and $B_y$ for a single snapshot and in panels (d) and (e)
the respective quantities after a $60$ minute average. Panel (c) shows the time averaged vertical component of the Poynting flux, whereas
panel (f) shows the Poynting flux computed from the $60$ minute averaged $\vec{v}$ and $\vec{B}$. While the time average strongly
reduces the amplitude of the velocity, many features present in the Poynting flux persist, indicating that there is a significant contribution
from velocity and magnetic field variations with long time scales. We further quantify the contributing time scales in Figure \ref{fig:16}, where we
plot the fraction of the Poynting flux remaining after application of the time averaging as function of the averaging time scale for the
QS (red), OF (green) and CA (blue) setups. The different line style correspond to different heights above the photosphere: $700$~km (solid),
$4.8$~Mm (dotted) and $8.9$~Mm (dashed). We use in the following discussion the ``median'' time scale at which $50\%$ of the Poynting flux is lost as characteristic time scale for the energy transport. For the height surface of $700$~km we find short time scales on the order of minutes, almost all of the energy flux is lost after averaging for more than a granular life time (10 minutes). This indicates that at this height the energy transport is dominated by granular motions braiding the magnetic field rooted in the intergranular lanes. As we move up, the dominant  time scales generally increase, but the amount of increase depends on the field geometry. While the OF setup shows time scales increasing to about 7 minutes, the CA setup shows time scales as long as 50 minutes. Magnetic field reaching to greater heights has in general a larger
footpoint separation, which naturally imposes longer convective time scales and these longer time scales have a significant contribution
to the energy transport. The OF case does not show an increase of time scale from $4.8$ to $8.9$ Mm since at these heights already the
imposed magnetic mean field dominates, i.e. the field is mostly vertical. In the case of the CA setup we find a small increase of the Poynting
flux for averaging time scales of a few minutes, potentially indicating wave excitation in the corona leading to a downward energy transport. 

Unfortunately we cannot apply the above analysis to the AR setup, since there active region decay is the dominant contribution to the Poynting flux, i.e.
the Poynting flux is actually negative and cannot be easily used for an analysis of coronal heating. However, extrapolating the trend towards longer time scales
we see from the QS and CA setup, we would expect even longer time scales in the AR setup.

For the closed flux QS and CA setups the time scales of 20 and 50 minutes in $8.9$~Mm height are significantly longer than typical
Alfv{\'e}nic time scales for propagation along magnetic field lines reaching into that height (about 4 minutes of the QS and less than a minute 
for the CA setup, assuming a length of about 25 Mm, and a mean Alfv{\'e}n velocity of about $100$ and $500$~km/s for the QS and CA setups, respectively).
The heating of the corona by (numerical) resistive and viscous heating is illustrated for the CA setup in Figure \ref{fig:17} as function of time (panel a) and 
space (panel b). The energy input is highly intermittent, we find typical energy releases on the order of $10^{19}$~erg~s$^{-1}$ per simulation pixel (peak values
can exceed $10^{21}$~erg~s$^{-1}$, the average heating per pixel is $1.7\times 10^{18}$~erg~s$^{-1}$ and 50\% of the heating comes from events with more than
$9\times 10^{18}$~erg~s$^{-1}$). Using auto-correlation in time and space and considering the
intervall where the auto-correlation is larger than $1/3$ we find an average lifetime of about $20$ and $80$ seconds  for $Q_{\rm res}$ and  $Q_{\rm visc}$, respectively.
The significantly shorter duration for $Q_{\rm res}$ is a consequence of the high magnetic Prandtl number we use in our setup.
The corresponding horizontal correlation lengths are about $500$ and $1000$~km. These values are averages over horizontal slices in $8.9$~Mm height. Using these
values we can estimate the energy release of a single heating event to be on the order of $10^{23}-10^{24}$~erg, which puts them close the $10^{24}$~erg
that were estimated by \citet{Parker:1988:nanoflare} for a nanoflare. A similar conclusion, based on a more detailed statistical analysis, was also presented
by \citet{Bingert:Peter:2013} for their corona simulations. This supports the view that the coronal heating for these closed field configurations is mostly of the ``DC'' type, 
i.e. braiding of field lines \citep{Parker:1972:topological_heating,Parker:1983:sheets} and subsequent energy release through nanoflares \citep{Parker:1988:nanoflare}.
This does not rule out the presence of ubiquitous waves with substantial amplitude (we find RMS velocities of about $30$~km~s$^{-1}$). In that respect our models agree
with the finding of \citet{Ballegooijen:2014:AC-DC} that braiding in the photosphere does excite waves with substantial amplitude. We find however, that this wave component 
does not directly contribute to the coronal energy transport, i.e. the Poynting flux is only moderately affected when these motions are filtered out. Because of that we
still classify this process as ``DC'' heating.

We cannot make easily a similar argument based on time scales for the OF setup (here 7 minutes might be still short enough compared to the potentially long
propagation speeds along open field lines to be considered ``AC'' or wave heating), but point out that the resistive and viscous heating is equally 
intermittent as in the CA case presented here, the corresponding correlation time scales are about a factor of two shorter, while spatial extent is comparable to the CA setup. 

\section{Computational efficiency}
\label{sec:efficiency}
One of our primary motivation for this investigation was enabling computationally inexpensive realistic corona simulations by using approximations for the two processes that pose the most stringent timestep constraints.
Compared to the photospheric version of the MURaM code the computational overhead in the coronal extension of our code is about $25\%$ (for one numerical timestep) and we demonstrated that coronal simulations 
reaching temperatures in the $1-5$~MK can be performed with numerical timesteps in the $0.05$ to $0.15$~seconds range for a the grid spacing of $192\times 192\times 64$~km$^3$ we considered here. In our treatment the effective speedup compared to the Alfv{\'e}nic timestep constraint is about $100$ in the AR case, while our speedup compared to the diffusive timestep constraint can exceed $10^3$. Evolving the AR setup for 1 hour of simulated time
requires about $5,000$ core hours if we use a value of $c=800$~km~s$^{-1}$.
 
\section{Conclusions}
\label{sec:concl}
We presented a new version of MURaM radiative MHD code that has been extended into the solar corona. For the corona we implemented optically thin radiative loss and field aligned heat conduction, while we treat
the ``chromosphere'' at this point in LTE. Similar to other so called ``realistic'' MHD simulations of the solar corona \citep[e.g.,][]{Gudiksen:etal:2011,Bingert:Peter:2011,Bingert:Peter:2013,Chen:etal:2014} we do not use
any parameterizations of coronal heating. Our setup includes the upper convection zone and photosphere where magneto-convection leads to the generation of a Poynting flux, which self-consistently heats the
upper layers of the simulation domain through a combination of (numerical) resistive and viscous energy dissipation.

Our implementation uses a fully explicit treatment and circumvents stringent time-step arising from the coronal Alfv{\'e}n velocity and heat conduction through the use of the ``Boris-correction'' \citep{Boris:1970:BC} and a 
hyperbolic treatment of heat conduction that imposes a maximum characteristic speed for conductive heat transport. Both approaches are inspired by semi-relativistic MHD as they are based on equations with a well
defined maximum propagation speed ``speed of light'', which is artificially reduced to a lower values for computational efficiency. 

We applied our code to four different coronal settings: quiet Sun, open flux, coronal arcade and active region and explored the dependence of the solutions on details of our adopted numerical diffusivity (process that heats the
corona in our simulation) and the chosen value for the peak Alfv{\'e}n as well as heat conduction speed ``reduced speed of light''. We tested  the latter using the active region setup in which the horizontally averaged
Alfv{\'e}n velocity reaches values of $10,000$ km~s$^{-1}$ and the peak value exceeds  $100,000$ km~s$^{-1}$. We explored values of $c$  from $400$ km~s$^{-1}$ to about $1600$ km~s$^{-1}$ and found no significant
influence on the resulting heating and mean temperature of the corona. It is remarkable that even values of $c$ similar to the speed of sound lead to reasonable results
when using the Boris correction as long as the flow velocity is artificially limited to assure the validity of the semi-relativistic approach. From our experiments we concluded that a setting of $c=\mbox{max}(C_S,\,3\,v)$ is a good compromise
between computational speed and sufficiently accurate treatment of MHD. This translates to values of $c$ in the $400-800$ km~s$^{-1}$ for the setups considered here. 
We found that our treatment of hyperbolic heat conduction was sufficiently accurate in all these cases except for features on the scale of the grid. Formally the solution of the hyperbolic heat conduction equation 
corresponds to a solution computed with a time averaged heat flux. Throughout most of the corona the associated averaging time scale is on the order of seconds,  i.e. short compared to typical time scales of interest. 

With respect to the treatment of numerical diffusivities we compared three setups differing in their effective numerical magnetic Prandtl number. While our reference setup has a high $P_{\rm m}$ (due to low magnetic diffusivity),
we considered also a $P_{\rm m}\sim1$ case with comparable diffusivities for velocity and magnetic field as well as a low  $P_{\rm m}$ setup with significantly reduced viscosity.
The most striking difference is found in the ratio of resistive to viscous heating, which is strongly reduced in a high magnetic Prandtl number setting. This is very similar to the behavior that was found by
\citet{Brandenburg:2011:SSD_low_Pm,Brandenburg:2014:Pm} for (high $\beta$) turbulence. While the ratio of resistive and viscous heating depends on numerical details, the sum of both terms (total coronal heating) is 
found to be very robust. The dependence on $P_{\rm m}$ ultimately illustrates that the microphysics are important if the goal is to determine how dissipation occurs. If the goal is just to quantify the total amount of energy dissipation, 
details of the dissipation process are less important. Furthermore, the fact that the Corona is essentially a very high $P_{\rm m}$ regime strongly suggests that resistive heating is negligible and dissipation happens through viscosity
on scales that are large enough to be captured in current numerical simulations.

We investigated four different magnetic field configurations that are representative of quiet Sun, open flux, coronal arcade and active regions. In all four setups we have in the convection zone part of the domain a small-scale dynamo 
operating that maintains a small-scale mixed polarity field. In the case of the quiet Sun setup the Poynting flux resulting from the small scale dynamo alone is sufficient to maintain an about 1 MK hot corona. Adding a small $3$~G
vertical mean field (open flux region) does not change the corona temperature significantly in spite of enhanced heating in the upper half of the simulation domain. The primary reason for that is more efficient cooling through heat conduction, which operates mostly in the vertical direction in this case. It is likely that the temperature would be lower if we would use more appropriate top boundary conditions (fully wave transmitting) in this case, but we did not investigate
that further. We find with about $2$~MK a significantly hotter corona in  the coronal arcade setup (a $\pm 50$ G vertical mean field in the left and right half of the domain). The active region setup is even hotter in the low corona (5-10 Mm above the photosphere) where we find hot loops with temperatures of up to $5$~MK, but the mean temperature drops significantly towards the top boundary, where it becomes comparable to the quiet Sun setup. The magnetic field
reaching into the upper parts of the simulation domain connects mostly to the umbrae of the spot pair in the photosphere where the Poynting flux is strongly suppressed. \citet{Chen:etal:2015NatPh,Cheung:etal:2015:DEM} found the strongest Poynting flux at the outer boundary of the umbra, which explains the hotter loops in the low corona for this setup.

We analyzed the time scales of motions that contribute to the coronal energy transport and found that for the closed field QS and CA setups 
the corresponding time scales are in the 20 to 50 minute range in a height of $8.9$~Mm above the photosphere. 
At the same time the viscous and resistive heating is highly intermittent with energy releases that are comparable to those expected in the nanoflare picture. Overall this supports 
the picture of ``DC'' heating by braiding of field lines on long time scales \citep{Parker:1972:topological_heating,Parker:1983:sheets}  and intermittent energy release in form of
nanoflares \citep{Parker:1988:nanoflare}. Similar to \citet{Ballegooijen:2014:AC-DC} we find that braiding in the photosphere does excite waves with substantial amplitudes of about 
$30$~km~s$^{-1}$, however, filtering that component out does only moderately impact the Poynting flux.

Overall we conclude that numerical ``tricks'' based on semi-relativistic MHD with an artificially reduced speed of light (Boris correction) enable a rather inexpensive modeling of the solar corona by effectively limiting both
Alfv{\'e}n and heat conduction speed. We did not find significant drawbacks from this approach in the setups considered here. The effects we found in the mean temperature from both artificial limitation of the Alfv{\'e}n velocity and
treatment of numerical diffusivity are comparable to those expected from the intrinsic uncertainty of the input physics (e.g. the assumed coronal element abundance affecting the radiative loss and the numerical value of the heat
conduction coefficient in the Spitzer formulation) as well as the intrinsic variability found in these simulations. The computational expense for our active region setup is about $5,000$ core hours for 1 hour of solar time. With the 
computing resources available today 3D realistic simulation of  the solar corona covering the full time span of active region formation to decay are well within reach.

\appendix
We present here a derivation of the semi-relativistic momentum equation following \citet{Gombosi:etal:2002:SR}. We start from the the MHD momentum equation of the form
\begin{equation}
	\varrho\frac{\partial \vec{v}}{\partial t}=-\varrho(\vec{v}\cdot\nabla)\vec{v}-\nabla p+\varrho\vec{g}+\frac{1}{c}\vec{j}\times\vec{B}\label{eq:app1}
\end{equation}
With the Maxwell equation 
\begin{equation}
	\nabla\times\vec{B}=\frac{4\pi}{c}\vec{j}+\frac{1}{c}\frac{\partial \vec{E}}{\partial t}
\end{equation}
we can rewrite the Lorentz force as
\begin{equation}
	\frac{1}{c}\vec{j}\times\vec{B}=\frac{1}{4\pi}(\nabla\times\vec{B})\times\vec{B}+\frac{1}{4\pi c}\vec{B}\times\frac{\partial \vec{E}}{\partial t}	
\end{equation}
With the relation $\vec{E}=-{1\over c} \vec{v}\times\vec{B}$ we can relate the time derivative of $\vec{E}$ to already known quantities:
\begin{equation}
	\frac{\partial \vec{E}}{\partial t}=-\frac{1}{c}\left(\frac{\partial \vec{v}}{\partial t}\times\vec{B} +\vec{v}\times\frac{\partial \vec{B}}{\partial t}\right)
\end{equation}
It is primarily the contribution from the first term that is responsible for limiting the Alfv{\'e}n velocity to values less than $c$ and we keep only that term in the following derivation. The second
term leads to additional forces perpendicular to the magnetic field that are important for an exact treatment of semi-relativistic MHD, but not required if we focus only on the reduction of Alfv{\'e}n velocity.
In addition it also follows that the second term is of order $v^2\over v_A^2$, i.e. small in the regime where semi-relativistic MHD is valid $v\ll c < v_A$: 
\begin{equation}
	\frac{\vert\vec{v}\times\frac{\partial \vec{B}}{\partial t}\vert}{\vert\frac{\partial \vec{v}}{\partial t}\times\vec{B}\vert}\sim\frac{\vert\vec{v}\times(\nabla\times(\vec{v}\times\vec{B}))\vert}{\vert{1\over 4\pi\varrho}((\nabla\times\vec{B})\times\vec{B})\times\vec{B}\vert}\sim\frac{v^2}{v_A^2}<\frac{v^2}{c^2}
\end{equation} 
Here $v_A=|\vec{B}|/\sqrt{4\pi\varrho}$ denotes the (classic) Alfv{\'e}n velocity. With this term the expression for the Lorentz force is given by:
\begin{eqnarray}
	\frac{1}{c}\vec{j}\times\vec{B}&=&\frac{1}{4\pi}(\nabla\times\vec{B})\times\vec{B}+\frac{1}{4\pi c^2}\vec{B}\times(\vec{B}\times\frac{\partial \vec{v}}{\partial t})\\
		&=&\frac{1}{4\pi}(\nabla\times\vec{B})\times\vec{B}-\frac{v_A^2}{c^2}[\mathcal{I}-\hat{\vec{b}}\hat{\vec{b}}]\varrho\frac{\partial \vec{v}}{\partial t}
\end{eqnarray}
where  $\hat{\vec{b}}=\vec{B}/|\vec{B}|$ denotes the unit vector in the direction of $\vec{B}$. Substituting this expression back into 
Eq. (\ref{eq:app1}) yields (using $x_A=v_A/c$):
\begin{equation}
	\left[\mathcal{I}+x_A^2(\mathcal{I}-\hat{\vec{b}}\hat{\vec{b}})\right]\varrho\frac{\partial \vec{v}}{\partial t}=-\varrho(\vec{v}\cdot\nabla)\vec{v}-\nabla p+\varrho\vec{g}+\frac{1}{4\pi}(\nabla\times\vec{B})\times\vec{B}
\end{equation}
The inverse of the ``enhanced inertia'' matrix on the left hand side is given by \citep{Gombosi:etal:2002:SR}:
 \begin{equation}
	\left[\mathcal{I}+x_A^2(\mathcal{I}-\hat{\vec{b}}\hat{\vec{b}})\right]^{-1}=\frac{1}{1+x_A^2}\left[\mathcal{I}+x_A^2\hat{\vec{b}}\hat{\vec{b}}\right]=
		\mathcal{I}-\frac{x_A^2}{1+x_A^2}\left[\mathcal{I}-\hat{\vec{b}}\hat{\vec{b}}\right]
\end{equation}
This leads to a momentum equation of the form:
\begin{equation}
	\frac{\partial \varrho\vec{v}}{\partial t}+\nabla\cdot(\varrho\vec{v}\vec{v}+\mathcal{I}p)=\varrho\vec{g}+\frac{1}{4\pi}(\nabla\times\vec{B})\times\vec{B}+\vec{F}_{SR}
\end{equation}	
where the ``semi-relativistic'' correction term is given by
\begin{equation}
	\vec{F}_{SR}=-\frac{x_A^2}{1+x_A^2}\left[\mathcal{I}-\hat{\vec{b}}\hat{\vec{b}}\right]\left(-\varrho(\vec{v}\cdot\nabla)\vec{v}-\nabla p+\varrho\vec{g}+\frac{1}{4\pi}(\nabla\times\vec{B})\times\vec{B}\right)\label{eq:FSR}
 \end{equation}	 
 Since our aim is not to compute an exact solution of semi-relativistic MHD, but rather to use the minimal amount of correction terms needed to limit the Alfv{\'e}n velocity, we can use some freedom in determining the
 quantity $x_A^2/(1+x_A^2)$ in front of the projection operator. We generalize this expression as $1-f_A$ and use
 \begin{equation}
 	f_A=\frac{1}{\sqrt{1+({v_A\over c})^4}}
 \end{equation}
 which leads to a limitation of the Alfv{\'e}n velocity in the following form:
 \begin{equation}
 	v_A^2 \longrightarrow \frac{v_A^2}{\sqrt{1+({v_A\over c})^4}}\;.\label{eq:va_lim}
 \end{equation}
 While we did not find that the detailed functional form of $f_A$ matters as long as the limited Alfv{\'e}n velocity  Eq. (\ref{eq:va_lim}) remains a monotonic function of $v_A$ and $f_A$ asymptotes as $ x_A^{-2}$ for large values of $x_A$,
 we prefer to use expressions for $f_A$ that have a sharper transition than the $(1+x_A^2)^{-1}$ that follows from semi-relativistic MHD in an attempt to minimize the volume of the simulation domain where the correction term  Eq. (\ref{eq:FSR})
 contributes.
 
 For solving the MHD equations we need to determine a maximum characteristic velocity that will be used for determining the time step as well as for computing numerical diffusivities. As shown by \citet{Gombosi:etal:2002:SR}
 the wave speeds in semi-relativistic MHD can be quite complicated. We use the following approximate expression:
 \begin{equation}
 	C_{\rm max}={\rm max}\left(C_S,\sqrt{f_A(C_S^2+v_A^2)}\right)+\vert\vec{v}\vert\label{Eq:Cmax}
 \end{equation}
 
\acknowledgements
The National Center for Atmospheric Research (NCAR) is sponsored by the National Science Foundation. The author thanks M.C.M Cheung and F. Chen for comments on the manuscript and the BIFROST team for providing equation of state tables. The author thanks the anonymous referee for suggestions that significantly improved the presentation of results. High-performance computing resources were provided by NCAR's Computational and Information Systems Laboratory, sponsored by the National Science Foundation, on Yellowstone (http://n2t.net/ark:/85065/d7wd3xhc) and by the NASA High-End Computing (HEC) Program through the NASA Advanced Supercomputing (NAS) Division at the Ames Research Center. This research has been partially supported through NASA contracts NNX14AI14G and NNX13AK54G.

\vspace{1cm}

\bibliographystyle{aasjournal}
\bibliography{natbib/papref_m}

\begin{thebibliography}{}
\expandafter\ifx\csname natexlab\endcsname\relax\def\natexlab#1{#1}\fi

\bibitem[{{Abbett}(2007)}]{Abbett:2007}
{Abbett}, W.~P. 2007, \apj, 665, 1469

\bibitem[{{Bingert} \& {Peter}(2011)}]{Bingert:Peter:2011}
{Bingert}, S., \& {Peter}, H. 2011, \aap, 530, A112

\bibitem[{{Bingert} \& {Peter}(2013)}]{Bingert:Peter:2013}
---. 2013, \aap, 550, A30

\bibitem[{{Blackman} \& {Field}(2003)}]{Blackman:Field:2003:non-fickian}
{Blackman}, E.~G., \& {Field}, G.~B. 2003, Physics of Fluids, 15, L73

\bibitem[{{Boris}(1970)}]{Boris:1970:BC}
{Boris}, J.~P. 1970, NRL Memorandum Report 2167

\bibitem[{{Bourdin} {et~al.}(2013){Bourdin}, {Bingert}, \&
  {Peter}}]{Bourdin:etal:2013:obs_driven}
{Bourdin}, P.-A., {Bingert}, S., \& {Peter}, H. 2013, \aap, 555, A123

\bibitem[{{Bradshaw} \& {Cargill}(2013)}]{Bradshaw:Cargill:2013:resolution}
{Bradshaw}, S.~J., \& {Cargill}, P.~J. 2013, \apj, 770, 12

\bibitem[{{Brandenburg}(2011)}]{Brandenburg:2011:SSD_low_Pm}
{Brandenburg}, A. 2011, \apj, 741, 92

\bibitem[{{Brandenburg}(2014)}]{Brandenburg:2014:Pm}
---. 2014, \apj, 791, 12

\bibitem[{{Brandenburg} {et~al.}(2004){Brandenburg}, {K{\"a}pyl{\"a}}, \&
  {Mohammed}}]{Brandenburg:etal:2004:non-fickian}
{Brandenburg}, A., {K{\"a}pyl{\"a}}, P.~J., \& {Mohammed}, A. 2004, Physics of
  Fluids, 16, 1020

\bibitem[{{Chen} \& {Peter}(2015)}]{Chen:Peter:2015}
{Chen}, F., \& {Peter}, H. 2015, \aap, 581, A137

\bibitem[{{Chen} {et~al.}(2014){Chen}, {Peter}, {Bingert}, \&
  {Cheung}}]{Chen:etal:2014}
{Chen}, F., {Peter}, H., {Bingert}, S., \& {Cheung}, M.~C.~M. 2014, \aap, 564,
  A12

\bibitem[{{Chen} {et~al.}(2015){Chen}, {Peter}, {Bingert}, \&
  {Cheung}}]{Chen:etal:2015NatPh}
---. 2015, Nature Physics, 11, 492

\bibitem[{{Cheung} {et~al.}(2015){Cheung}, {Boerner}, {Schrijver}, {Testa},
  {Chen}, {Peter}, \& {Malanushenko}}]{Cheung:etal:2015:DEM}
{Cheung}, M.~C.~M., {Boerner}, P., {Schrijver}, C.~J., {et~al.} 2015, \apj,
  807, 143

\bibitem[{{Dedner} {et~al.}(2002){Dedner}, {Kemm}, {Kr{\"o}ner}, {Munz},
  {Schnitzer}, \& {Wesenberg}}]{Dedner:etal:2002:divB}
{Dedner}, A., {Kemm}, F., {Kr{\"o}ner}, D., {et~al.} 2002, Journal of
  Computational Physics, 175, 645

\bibitem[{{Fisher} {et~al.}(1985){Fisher}, {Canfield}, \&
  {McClymont}}]{Fisher:etal:1985:flare}
{Fisher}, G.~H., {Canfield}, R.~C., \& {McClymont}, A.~N. 1985, \apj, 289, 414

\bibitem[{{Gombosi} {et~al.}(1993){Gombosi}, {Jokipii}, {Kota}, {Lorencz}, \&
  {Williams}}]{Gombosi:1993:telegraph}
{Gombosi}, T.~I., {Jokipii}, J.~R., {Kota}, J., {Lorencz}, K., \& {Williams},
  L.~L. 1993, \apj, 403, 377

\bibitem[{{Gombosi} {et~al.}(2002){Gombosi}, {T{\'o}th}, {De Zeeuw}, {Hansen},
  {Kabin}, \& {Powell}}]{Gombosi:etal:2002:SR}
{Gombosi}, T.~I., {T{\'o}th}, G., {De Zeeuw}, D.~L., {et~al.} 2002, Journal of
  Computational Physics, 177, 176

\bibitem[{{Gudiksen} {et~al.}(2011){Gudiksen}, {Carlsson}, {Hansteen}, {Hayek},
  {Leenaarts}, \& {Mart{\'{\i}}nez-Sykora}}]{Gudiksen:etal:2011}
{Gudiksen}, B.~V., {Carlsson}, M., {Hansteen}, V.~H., {et~al.} 2011, \aap, 531,
  A154

\bibitem[{{Gudiksen} \& {Nordlund}(2002)}]{Gudiksen:Nordlund:2002}
{Gudiksen}, B.~V., \& {Nordlund}, {\AA}. 2002, \apj, 572, L113

\bibitem[{{Gudiksen} \&
  {Nordlund}(2005{\natexlab{a}})}]{Gudiksen:Nordlund:2005b}
---. 2005{\natexlab{a}}, \apj, 618, 1031

\bibitem[{{Gudiksen} \&
  {Nordlund}(2005{\natexlab{b}})}]{Gudiksen:Nordlund:2005a}
---. 2005{\natexlab{b}}, \apj, 618, 1020

\bibitem[{{Guidoni} \& {Longcope}(2010)}]{Guidoni:Longcope:2010}
{Guidoni}, S.~E., \& {Longcope}, D.~W. 2010, \apj, 718, 1476

\bibitem[{{Gustafsson} {et~al.}(1975){Gustafsson}, {Bell}, {Eriksson}, \&
  {Nordlund}}]{Gustafsson:etal:1975}
{Gustafsson}, B., {Bell}, R.~A., {Eriksson}, K., \& {Nordlund}, A. 1975, \aap,
  42, 407

\bibitem[{{Klimchuk}(2006)}]{Klimchuk:2006:review}
{Klimchuk}, J.~A. 2006, \solphys, 234, 41

\bibitem[{{Landi} {et~al.}(2012){Landi}, {Del Zanna}, {Young}, {Dere}, \&
  {Mason}}]{Landi:2012:Chianti7}
{Landi}, E., {Del Zanna}, G., {Young}, P.~R., {Dere}, K.~P., \& {Mason}, H.~E.
  2012, \apj, 744, 99

\bibitem[{{Lie-Svendsen} {et~al.}(2001){Lie-Svendsen}, {Leer}, \&
  {Hansteen}}]{Lie-Svendsen:etal:2001}
{Lie-Svendsen}, {\O}., {Leer}, E., \& {Hansteen}, V.~H. 2001, \jgr, 106, 8217

\bibitem[{{Longcope} {et~al.}(2009){Longcope}, {Guidoni}, \&
  {Linton}}]{Longcope:etal:2009}
{Longcope}, D.~W., {Guidoni}, S.~E., \& {Linton}, M.~G. 2009, \apjl, 690, L18

\bibitem[{{Lyon} {et~al.}(2004){Lyon}, {Fedder}, \&
  {Mobarry}}]{Lyon:etal:2004:LFM}
{Lyon}, J.~G., {Fedder}, J.~A., \& {Mobarry}, C.~M. 2004, Journal of
  Atmospheric and Solar-Terrestrial Physics, 66, 1333

\bibitem[{{Meyer} {et~al.}(2012){Meyer}, {Balsara}, \&
  {Aslam}}]{Meyer:2012:super-timestepping}
{Meyer}, C.~D., {Balsara}, D.~S., \& {Aslam}, T.~D. 2012, \mnras, 422, 2102

\bibitem[{{Mok} {et~al.}(2005){Mok}, {Miki{\'c}}, {Lionello}, \&
  {Linker}}]{Mok:etal:2005}
{Mok}, Y., {Miki{\'c}}, Z., {Lionello}, R., \& {Linker}, J.~A. 2005, \apj, 621,
  1098

\bibitem[{{Mok} {et~al.}(2008){Mok}, {Miki{\'c}}, {Lionello}, \&
  {Linker}}]{Mok:etal:2008}
---. 2008, \apjl, 679, L161

\bibitem[{{Parker}(1972)}]{Parker:1972:topological_heating}
{Parker}, E.~N. 1972, \apj, 174, 499

\bibitem[{{Parker}(1983)}]{Parker:1983:sheets}
---. 1983, \apj, 264, 642

\bibitem[{{Parker}(1988)}]{Parker:1988:nanoflare}
---. 1988, \apj, 330, 474

\bibitem[{{Peter} {et~al.}(2004){Peter}, {Gudiksen}, \&
  {Nordlund}}]{Peter:etal:2004}
{Peter}, H., {Gudiksen}, B.~V., \& {Nordlund}, {\AA}. 2004, \apjl, 617, L85

\bibitem[{{Peter} {et~al.}(2006){Peter}, {Gudiksen}, \&
  {Nordlund}}]{Peter:etal:2006}
---. 2006, \apj, 638, 1086

\bibitem[{{Rempel}(2014)}]{Rempel:2014:SSD}
{Rempel}, M. 2014, \apj, 789, 132

\bibitem[{{Rempel}(2015)}]{Rempel:2015:moat}
---. 2015, \apj, 814, 125

\bibitem[{{Rempel} {et~al.}(2009){Rempel}, {Sch{\"u}ssler}, \&
  {Kn{\"o}lker}}]{Rempel:etal:2009}
{Rempel}, M., {Sch{\"u}ssler}, M., \& {Kn{\"o}lker}, M. 2009, \apj, 691, 640

\bibitem[{{Rogers} {et~al.}(1996){Rogers}, {Swenson}, \&
  {Iglesias}}]{Rogers:opal:1996}
{Rogers}, F.~J., {Swenson}, F.~J., \& {Iglesias}, C.~A. 1996, \apj, 456, 902

\bibitem[{{Schunk}(1975)}]{Schunk:1975}
{Schunk}, R.~W. 1975, \planss, 23, 437

\bibitem[{{Snodin} {et~al.}(2006){Snodin}, {Brandenburg}, {Mee}, \&
  {Shukurov}}]{Snodin:etal:2006:non-fickian}
{Snodin}, A.~P., {Brandenburg}, A., {Mee}, A.~J., \& {Shukurov}, A. 2006,
  \mnras, 373, 643

\bibitem[{{Spitzer}(1962)}]{Spitzer:1962}
{Spitzer}, L. 1962, Physics of Fully Ionized Gases, Interscience Publishers,
  New York (Interscience Publishers, New York)

\bibitem[{{V{\" o}gler} {et~al.}(2005){V{\" o}gler}, {Shelyag}, {Sch{\"
  u}ssler}, {Cattaneo}, {Emonet}, \& {Linde}}]{Voegler:etal:2005}
{V{\" o}gler}, A., {Shelyag}, S., {Sch{\" u}ssler}, M., {et~al.} 2005, \aap,
  429, 335

\bibitem[{{van Ballegooijen} {et~al.}(2014){van Ballegooijen}, {Asgari-Targhi},
  \& {Berger}}]{Ballegooijen:2014:AC-DC}
{van Ballegooijen}, A.~A., {Asgari-Targhi}, M., \& {Berger}, M.~A. 2014, \apj,
  787, 87

\bibitem[{{van der Holst} {et~al.}(2014){van der Holst}, {Sokolov}, {Meng},
  {Jin}, {Manchester}, {T{\'o}th}, \& {Gombosi}}]{vdHolst:etal:2014:AWSoM}
{van der Holst}, B., {Sokolov}, I.~V., {Meng}, X., {et~al.} 2014, \apj, 782, 81

\bibitem[{{Withbroe} \& {Noyes}(1977)}]{Withbroe:Noyes:1977:corona_energy}
{Withbroe}, G.~L., \& {Noyes}, R.~W. 1977, \araa, 15, 363

\end{thebibliography}

\end{document}